\documentclass[longauth]{aa}

\usepackage[utf8]{inputenc}
\usepackage{graphicx}

\usepackage{amssymb}
\usepackage{amsmath}
\usepackage{float}
\usepackage{natbib}
\bibpunct{(}{)}{;}{a}{}{,} 
\usepackage{array}
\usepackage{hyperref}
\usepackage{subfig}
\usepackage[varg]{txfonts}
\usepackage{multirow}
\usepackage{siunitx}
\usepackage{upgreek}
\usepackage{textgreek}
\usepackage{ulem}
\usepackage{titlesec}
\usepackage{tablefootnote}
\usepackage{lscape}
\usepackage{xcolor}
\usepackage{makecell}
\usepackage{adjustbox}

\hypersetup{colorlinks=true,citecolor=blue}

\titleformat{\paragraph}
{\normalfont\normalsize\bfseries}{\theparagraph}{1em}{}
\titlespacing*{\paragraph}
{0pt}{3.25ex plus 1ex minus .2ex}{1.5ex plus .2ex}

\def\app#1#2{%
  \mathrel{%
    \setbox0=\hbox{$#1\sim$}%
    \setbox2=\hbox{%
      \rlap{\hbox{$#1\propto$}}%
      \lower1.1\ht0\box0%
    }%
    \raise0.25\ht2\box2%
  }%
}

\DeclareTextSymbol{\degre}{T1}{6}
\DeclareTextSymbol{\degre}{OT1}{23}

\makeatletter
\g@addto@macro{\endtabular}{\rowfont{}}
\makeatother
\newcommand{\rowfonttype}{}
\newcommand{\rowfont}[1]{
\gdef\rowfonttype{#1}#1\ignorespaces%
}
\makeatother

\defcitealias{}{}          

\begin{document}

  \title{Automatic model-based telluric correction for the ESPRESSO data reduction software\thanks{Based on guaranteed time observations collected at the European Southern Observatory under ESO program 1102.C-0744 by the ESPRESSO Consortium.}}
  \subtitle{Model description and application to radial velocity computation}

   \author{ R. Allart\inst{1,2,*,\thanks{Trottier Postdoctoral Fellow}} ,
   C. Lovis\inst{2},
   J. Faria\inst{3,4},
   X. Dumusque\inst{2},
   D. Sosnowska\inst{2},
   P. Figueira\inst{2,3},
   A. M. Silva\inst{3,4},
   A. Mehner\inst{5},
   F. Pepe\inst{2},
   S. Cristiani\inst{6,7},
   R. Rebolo\inst{8,9,10},
   N. C. Santos\inst{3,4},
   V. Adibekyan\inst{3,4},
   G. Cupani\inst{6,7},
   P. Di Marcantonio\inst{6},
   V. D'Odorico\inst{5,6},
   J. I. Gonz\'alez Hern\'andez\inst{8,9},
   C. J. A. P. Martins\inst{3,11},
   D. Milakovi\'c\inst{7,6,12},
   N.  J. Nunes\inst{13},
   A. Sozzetti\inst{14},
   A. Su\'arez Mascare\~no\inst{8,9},
   H. Tabernero\inst{15} and
   M. R. Zapatero Osorio\inst{15}}
   
   \institute{
   \inst{1} Department of Physics, and Institute for Research on Exoplanets, Universit\'e de Montr\'eal, Montr\'eal, H3T 1J4, Canada\\
   \inst{2} Observatoire astronomique de l'Universit\'e de Gen\`eve, Universit\'e de Gen\`eve, chemin Pegasi 51, CH-1290 Versoix, Switzerland\\
   \inst{3} Instituto de Astrof\'{\i}sica e Ci\^encias do Espa\c co, CAUP, Universidade do Porto, Rua das Estrelas, 4150-762, Porto, Portugal \\
   \inst{4} Departamento de F\'{\i}sica e Astronomia, Faculdade de Ci\^encias, Universidade do Porto, Rua Campo Alegre, 4169-007, Porto, Portugal\\
   \inst{5} ESO, European Southern Observatory, Alonso de Cordova 3107, Vitacura, Santiago\\
   \inst{6} INAF – Osservatorio Astronomico di Trieste, via Tiepolo 11, 34143 Trieste, Italy\\
   \inst{7} Institute for Fundamental Physics of the Universe, IFPU, Via Beirut 2, 34151 Grignano, Trieste, Italy\\
   \inst{8} Instituto de Astrof{\'\i}sica de Canarias, V{\'\i}a L\'actea, 38205 La Laguna, Tenerife, Spain\\
   \inst{9} Universidad de La Laguna, Departamento de Astrof{\'\i}sica, 38206 La Laguna, Tenerife, Spain\\
   \inst{10} Consejo Superior de Investigaciones Científicas, E-28006 Madrid, Spain\\   
   \inst{11} Centro de Astrof\'{\i}sica da Universidade do Porto, Rua das Estrelas, 4150-762 Porto, Portugal\\
   \inst{12} INFN, Sezione di Trieste, Via Valerio 2, 34127 Trieste, Italy\\
   \inst{13} Instituto de Astrof\'{\i}sica e Ci\^encias do Espa\c co, Faculdade da Universidade de Lisboa, Edifício C8, Campo Grande, PT1749-016\\  
   \inst{14} INAF - Osservatorio Astrofisico di Torino, via Osservatorio 20, 10025 Pino Torinese, Italy\\
   \inst{15} Centro de Astrobiolog\'\i a (CSIC-INTA), Crta. Ajalvir km 4, E-28850 Torrej\'on de Ardoz, Madrid, Spain\\            
                        * \email{romain.allart@umontreal.ca}
             }

   \date{Received January 1, 2015; accepted January 1, 2015}


  \abstract
    {Ground-based high-resolution spectrographs are key instruments for several astrophysical domains, such as exoplanet studies. Unfortunately, the observed spectra are contaminated by the Earth's atmosphere and its large molecular absorption bands. While different techniques (forward radiative transfer models,  principle component analysis (PCA), or other empirical methods) exist to correct for telluric lines in exoplanet atmospheric studies, in radial velocity (RV) studies, telluric lines with an absorption depth of $>$2\,\%  are generally masked, which poses a problem for faint targets and M dwarfs as most of their RV content is present where telluric contamination is important.}
   {We propose a simple telluric model to be embedded in the Echelle SPectrograph for Rocky Exoplanets and Stable Spectroscopic Observations (ESPRESSO) data reduction software (DRS). The goal is to provide telluric-free spectra and enable RV measurements through the cross-correlation function technique (and others), including spectral ranges where telluric lines fall.}
    {The model is a line-by-line radiative transfer code that assumes a single atmospheric layer. We use the sky conditions and the physical properties of the lines from the HITRAN database to create the telluric spectrum. This high-resolution model is then convolved with the instrumental resolution and sampled to the instrumental wavelength grid. A subset of selected telluric lines is used to robustly fit the spectrum through a Levenberg-Marquardt minimization algorithm.}
   {We computed the model to the H$_2$O lines in the spectral range of ESPRESSO. When applied to stellar spectra from A0- to M5-type stars, the residuals of the strongest water lines are below the 2\% peak-to-valley (P2V) amplitude for all spectral types, with the exception of M dwarfs, which are within the pseudo-continuum. We then determined the RVs from the telluric-corrected ESPRESSO spectra of Tau Ceti and Proxima. We created telluric-free masks and compared the obtained RVs with the DRS RVs. In the case of Tau Ceti, we identified that micro-telluric lines introduce systematics up to an amplitude of 58\,$\mathrm{cm \cdot s^{-1}}$  and with a period of one year if not corrected. For Proxima, the impact of micro-telluric lines is negligible due to the low flux below 5900\,\AA. For late-type stars, the gain in spectral content at redder wavelengths is equivalent to a gain of 25\% in photon noise or a factor of 1.78 in exposure time. This leads to better constraints on the semi-amplitude and eccentricity of Proxima d, which was recently proposed as a planet candidate. Finally, we applied our telluric model to the O$_2$ $\gamma$-band and we obtained residuals below the 2\% P2V amplitude.}
 {We propose a simple telluric model for high-resolution spectrographs to correct individual spectra and to achieve precise RVs. The removal of micro-telluric lines, coupled with the gain in spectral range, leads to more precise RVs. Moreover, we showcase that our model can be applied to other molecules, and thus to other wavelength regions observed by other spectrographs, such as NIRPS.}

   \titlerunning{Telluric correction}
   \authorrunning{ R. Allart, C. Lovis J. Faria et al. }
   \maketitle
%
\section{Introduction}

The Earth's atmosphere is rich in spectral features. The molecules H$_2$O, O$_2$, O$_3$, CH$_4$, and CO$_2$ are the main molecules from 0.3 to 5\,microns in the Earth's transmission spectrum (also called the telluric spectrum). When astrophysical objects are observed with ground-based instruments, their light is absorbed by the Earth's atmosphere, and thus their spectrum is contaminated by the telluric spectrum. At high spectral resolution, the telluric lines are well resolved and thousands of them can be distinguished. Moreover, at specific wavelengths, they completely absorb the light of the observed object, thus losing any astrophysical content. The absorption of the telluric spectrum varies as a function of the volume mixing ratio along the line of sight, which is controlled at first order by airmass. The exception for this is the water vapor, for which water column density and total content in the line of sight changes inhomogeneously across the sky and over a timescale of hours due to the weather conditions \citep{li_temporal_2018}.  \\
\\
With increasing instrumental performances and a larger telescope collecting area, developing methodologies to measure high-fidelity unbiased spectra is becoming a real challenge for several astrophysical topics such as an accurate stellar library, stellar activity indicators, precise stellar radial velocity (RV), exoplanet atmospheric characterization, or the measurement of the fine structure constant. Therefore, accurate correction of the Earth's transmission spectrum is crucial in the process of obtaining high-fidelity results. Several methods are available for removing the telluric spectrum from complex model-based algorithms (e.g., \texttt{Molecfit} \citep{smette_molecfit_2015, kausch_molecfit_2015}, \texttt{TAPAS} \citep{bertaux_tapas_2014}, and \texttt{TELFIT} \citep{gullikson_correcting_2014}) to the observation of a spectrophotometric standard star close in time and in the sky to the scientific observation \citep{vidal-madjar_deuterium_1986}. \cite{ulmer-moll_telluric_2019} compared these different approaches for near-infrared archival data obtained with CRIRES. The authors concluded that the model-based approach better corrects water telluric lines while the standard star is more appropriate for correcting the O$_2$ lines. Such methods can be applied to different astrophysical objects. Other studies are more focused on the detection of exoplanets through RV and on the characterization of their atmospheres. For example, \cite{artigau_telluric-line_2014} proposed an approach based on principal component analysis (PCA) built over a library of spectrophotometric standard stars for archival HARPS data. They obtained a better RV accuracy by including a wider telluric-corrected spectral domain when compared with the restricted telluric-free domain used. Empirical approaches, of which PCA is one (e.g., SYSREM; \cite{mazeh_sys-rem_2007}), are also often used for exoplanet atmospheric studies and take advantage of the different RV components of the planet, the star, and the Earth \citep{snellen_orbital_2010}. In addition to PCA-based algorithms, forward modeling algorithms are used to study exoplanet atmospheres \citep[e.g.,][]{allart_search_2017}. However, forward modeling algorithms are not yet optimized to automatically produce precise RVs as their minimization algorithm can be biased by stellar contamination depending on the combination of the spectral type, systemic velocity, and barycentric Earth radial velocity (BERV). \cite{allart_search_2017} demonstrated this with \texttt{Molecfit}, which is very sensitive to the selection of telluric regions that are to be fitted, in order to perform the most optimal correction.\\
\\
Precise RVs are obtained, with stable high-resolution spectrographs, by cross-correlating stellar spectra with a binary mask \citep{baranne_elodie_1996,pepe_coralie_2002} or through template-matching algorithms \citep[e.g.,][]{anglada-escude_harps-terra_2012,astudillo-defru_harps_2015,zechmeister_spectrum_2018}. Telluric lines spoil RV measurements when they are not removed or corrected for, independently of the method used to measure RVs. Exclusion of the telluric lines can be quickly done, either by masking the complete bands or by masking each line manually on each spectrum \citep[e.g.,][]{astudillo-defru_harps_2015}. Only the telluric lines of O$_2$, H$_2$O, and OH affect measurements in the visible wavelength. Since these cover a comparatively small fraction of the wavelength range, discarding them from the analysis has a comparatively small effect. However, with the development of high-resolution spectrographs in the near-infrared (CARMENES \citep{quirrenbach_carmenes_2014}, SPIRou \citep{donati_spirou_2020}, and NIRPS \citep{wildi_nirps_2017}) or installed on larger collecting areas (ESPRESSO \citep{pepe_espresso_2021}), the need for an automatic and homogeneous telluric removal tool became apparent. Moreover, \cite{cunha_impact_2013}, \cite{lisogorskyi_activity_2019}, and \cite{xuesong_wang_effects_2019} noticed that micro-telluric lines (unresolved or very shallow lines, with a depth up to 2\%), which are generally not excluded when building the cross-correlation function (CCF), can introduce systematics in the RV at a level of 10\,-\,20\,$ \mathrm{cm \cdot s^{-1}}$, which is comparable to the expected RV precision of the state-of-the-art optical spectrographs (ESPRESSO \citep{pepe_espresso_2021}, EXPRES \citep{jurgenson_expres_2016}) and to the precision required in order to detect an Earth-Sun system analog.\\
\\
In recent years, several techniques have been developed to overcome the limitation of (micro-) telluric lines to reach extremely precise RVs. Similar to \cite{artigau_telluric-line_2014}, \cite{leet_toward_2019} propose the self-calibrating, empirical, light-weight linear regression telluric (SELENITE) model. They use the variation in water telluric line depths as a function of the precipitable water vapor and the variation in non-water lines with the airmass to model the telluric spectrum. They do not require any line list database, but instead they use the observations of a few dozen B stars to build their SELENITE telluric database. The authors applied their correction to EXPRES data and reached average residuals similar to model-based corrections discussed in \cite{ulmer-moll_telluric_2019}. \cite{artigau_toi-1278_2021} propose a two-step approach where they apply the first correction with a \texttt{TAPAS} model scaled to the water content and airmass. Then, they use the PCA-based approach discussed in \cite{artigau_telluric-line_2014} on the residuals of the \texttt{TAPAS} correction. The authors applied their technique to SPIRou data (in the near-infrared, which is more polluted by telluric lines in comparison to the visible) of TOI-1278. Finally, \cite{bedell_wobble_2019} and \cite{cretignier_yarara_2021} also propose data-driven approaches with the Wobble and YARARA algorithms that use the time series of a given star to disentangle the contribution of the telluric spectrum from the stellar spectrum. Therefore, their methods can only be applied as a postprocessing step as they require a large number of spectra of the same object with a large enough variation in the BERV. With additional analysis treatment, the authors are able to reduce the RV root mean square (rms) by 10-20\% in comparison to the classical reduction for HARPS observations.

\begin{figure*}[t]
\resizebox{\hsize}{!}{\includegraphics{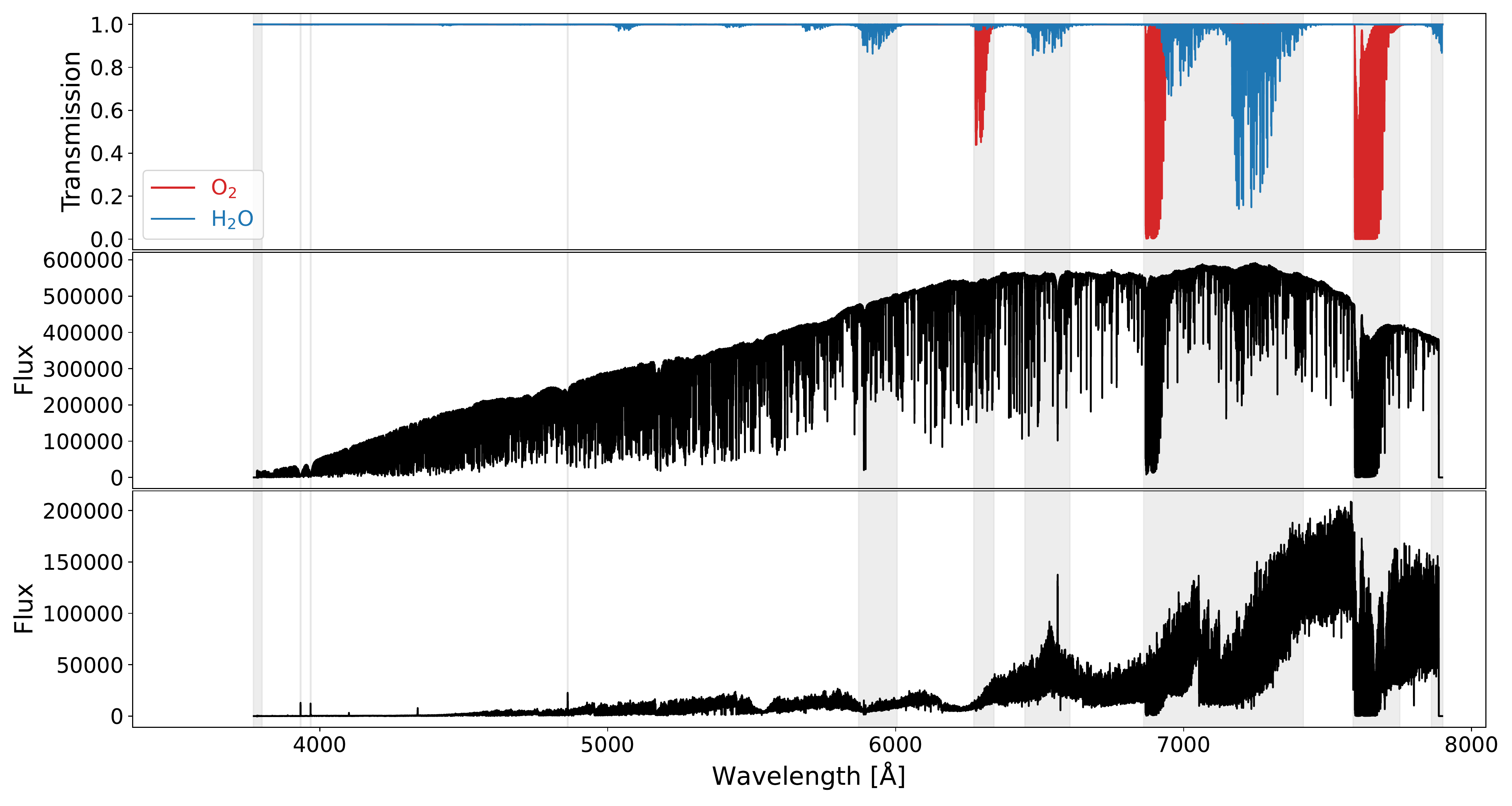}}
\caption{Telluric and stellar spectra over the ESPRESSO spectral range. The gray bands are the regions excluded from the mask used in the ESPRESSO DRS. \textit{Top panel:} Telluric spectrum produced with \texttt{Molecfit} \citep{smette_molecfit_2015, kausch_molecfit_2015}. The H$_2$O lines are in blue, while the O$_2$ lines are in red. \textit{Middle panel:} ESPRESSO spectrum of the G-type star Tau Ceti. \textit{Bottom panel:} ESPRESSO spectrum of the M5-type star Proxima.}
\label{ESPRESSO_tell}
\end{figure*}
Figure\,\ref{ESPRESSO_tell} shows where the telluric lines fall on top of G- and M5-type stars observed with the ESPRESSO spectrograph. Both H$_2$O and O$_2$ bands are present across the full spectral range of ESPRESSO but their intensities increase toward redder wavelengths. The ESPRESSO data reduction software (DRS), which builds upon the HARPS DRS, excludes the wavelength ranges (gray bands) of the telluric lines and the active stellar lines (Ca\,\ion{II}\ H \& K, Na doublet, and Balmer lines)  when the CCFs are built. This drastic measure masks up to $\sim$27\%  of the spectral range of ESPRESSO, which is essential for analyzing late-type stars as their blackbody maxima are toward redder wavelengths. Moreover, lines at redder wavelengths are known to be less impacted by stellar activity \citep{suarez_mascareno_revisiting_2020}. Micro-telluric water bands are not excluded when the CCFs are built (5000-5900\,\AA). The impact of these lines on the RV content of HARPS was considered negligible in comparison to the number of stellar lines present at the same positions. However, the collecting area of the Very Large Telescope (VLT), the wider spectral range to redder wavelengths, and the improved instrumental stability of ESPRESSO increase their impact, arguably to a detectable level, as demonstrated by \cite{cunha_impact_2013}. \\

The aforementioned telluric correction techniques are not practical for integration into an automatic pipeline to reduce individual frames and to be used by the whole community. In addition, having an additional database of telluric standards and complex forward models that require manual intervention are clear limitations. Therefore, we explore in this paper how a simple telluric model could rapidly correct the spectra of astrophysical objects with a minimal number of assumptions. The telluric model is based on a line-by-line radiative transfer code that assumes a single atmospheric layer. It only requires information on the sky observations, knowledge of the instrumental resolution, and a line list database such as HITRAN \citep{rothman_hitran2012_2013}. This algorithm will be implemented in the official DRS of ESPRESSO. Therefore, the model is built for the \texttt{S2D\_BLAZE\_A} spectra of ESPRESSO from which the CCFs are derived. The \texttt{S2D\_BLAZE\_A} product is the echelle-extracted spectra in the Earth's rest frame produced by the ESPRESSO DRS before removing the blaze to preserve the relative lines' strengths. We present a special focus on the telluric correction of water lines, which are the most numerous and variable, and therefore the most problematic telluric lines in the spectral range of ESPRESSO.\\

In Sect.\,\ref{Sec_model_description} we describe our one-layer telluric model. In Sect.\,\ref{Sec_model_fit}, we describe how the telluric model parameters are adjusted to the data. In Sect.\,\ref{Sec_App_Stellar_type}, we apply the telluric correction to different spectral types. Section\,\ref{Sec_App_RV} presents the implications of the telluric correction of water for the RV time series of Tau Ceti and Proxima obtained with ESPRESSO.  Afterward, we open the discussion to telluric correction of the O$_2$ molecular bands with its implication for other molecules and spectral domains in Sect.\,\ref{Sec_O2}. We conclude in Sect.\,\ref{Sec_concl}.
\section{Model description}\label{Sec_model_description}

To build the Earth's transmission spectrum, one needs to consider that its atmosphere is a gas that attenuates the stellar light as a function of the optical path length. Following the description on the HITRAN website\footnote{\url{https://hitran.org/docs/definitions-and-units/}}, the telluric spectrum $T(\nu)$ follows the Beer-Lambert law and can be expressed as a function of the sum of the opacity function $\tau(\nu)$ of each telluric line. The telluric spectrum is built in wavenumber units, $\nu$, in $\mathrm{cm^{-1}}$: 

\begin{eqnarray}
T(\nu) = e^{- \sum \tau(\nu) }
\label{T_nu}.
\end{eqnarray}  

The opacity function of a single line depends on the absorption cross-section $\sigma(\nu)$  in $\mathrm{cm^{2} / molecule}$, the number density $n(z)$ in $\mathrm{molecule \cdot cm^{-3}}$, and the path length $l$ in centimeters:

\begin{eqnarray}
\tau(\nu) = \int_{0}^{l}\sigma(\nu) \cdot n(z) \cdot dz
\label{tau_nu}.
\end{eqnarray}  

If a single atmospheric layer at constant pressure and a homogeneous distribution of water vapor is considered, we can see that
\begin{eqnarray} 
\int_{0}^{l} n(z)\cdot dz = n\cdot l,
\end{eqnarray}  
and
\begin{eqnarray} 
IWV_{LOS} = \dfrac{n\cdot l}{n_{H_{2}O}},
\end{eqnarray}   
where $n_{H_{2}O} = 3.34 \cdot 10^{22}$ $\mathrm{molecules \cdot cm^{-3}}$, is a constant expressing the number of water molecules per unit volume of liquid water, and $IWV_{LOS}$ is the integrated water vapor in the line of sight, which represents the condensed water column in millimeters.

Therefore, Equation\,\ref{T_nu} can be rewritten as

\begin{eqnarray}
T(\nu) = e^{- IWV_{LOS} \cdot n_{H_2O} \cdot \sum \sigma(\nu)  }
\label{T_nu_const}.
\end{eqnarray}  

The absorption cross-section of each line can be expressed as the line intensity $S_{ij}^{*}$ times the line profile, $f(\nu)$:

\begin{eqnarray}
\sigma(\nu) = S_{ij}^{*} \cdot f(\nu)
\label{sigma_nu},
\end{eqnarray}where $S_{ij}^{*}$ is the line intensity at a given temperature, $T$, and is linked to the reference line intensity $S_{ij}$ at ambient temperature (296\,K):

\begin{eqnarray}
S_{ij}^{*}=S_{ij} \cdot \dfrac{Q_{296K}}{Q_T} \cdot \dfrac{exp(-c_2 \cdot E" / T )}{exp(-c_2 \cdot E" / 296) } \cdot \dfrac{1 - exp(-c_2 \cdot \nu_{ij} / T )}{1 - exp(-c_2 \cdot \nu_{ij} / 296) }
\label{S_ij}.
\end{eqnarray}

Similarly, the line positions, $\nu_{ij}^{*}$, are subject to pressure variations, where $ \delta_{air}$ is the pressure shift and $P$ is the gas pressure:
                                
\begin{eqnarray}
\nu_{ij}^{*}=\nu_{ij} + \delta_{air} \cdot P
\label{nu_ij}.
\end{eqnarray}
To build a telluric spectrum that we will be able to use for the echelle-extracted spectra produced by the ESPRESSO DRS (S2D hereinafter), we build the model by spectral order. The wavenumber grid has a resolution of 3\,000\,000 and the grid is larger than the S2D grid by 40\,$ \mathrm{km \cdot s^{-1}}$ (1\,\AA\ at 7500\,\AA)  on each side of the orders to account for the BERV that shifts the spectrum on the detectors by, at most, 30\,$\mathrm{km \cdot s^{-1}}$. The wavenumber grid is then converted into angstrom and convolved with a Gaussian instrumental profile to produce the spectrum as would be seen by ESPRESSO. The width of the convolving Gaussian corresponds to the width of the Gaussian on the previously produced spectral resolution map (discussed in Sect\,\ref{subsection_resolution}). Finally, the last step consists in resampling the convolved telluric model to the S2D grid.

\subsection{Line profiles}
Different line profiles to model the telluric lines have been considered: Gaussian, Lorentzian, and Voigt profiles.

\subsubsection{Gaussian profile}
The Gaussian profile is the theoretical profile describing the thermal broadening of a line and is expressed as 
\begin{eqnarray}
f_{Gaussian}(\nu) = \sqrt{\dfrac{\mathrm{ln}(2)}{\pi\cdot \alpha^2}}\cdot \mathrm{exp}\left(-\dfrac{(\nu -\nu_{ij}^{*} )^2\cdot\mathrm{ln}(2)}{\alpha^2}\right)
\label{Gprofile},
\end{eqnarray}
 where $\alpha$ is the Gaussian half width at half maximum (HWHM) that depends on the Avogadro number $N_A\,=\,6.02\cdot10^{23}$\,mol$^{-1}$, the Boltzmann constant $k_{b} = 1.38\cdot10^{-23}\,\mathrm{kg\cdot m^2\cdot s^{-2}}$, the gas temperature $T$, the speed of light $c$, and the molar mass of water, $M_{mol} = 18.01\,\mathrm{g\cdot mol^{-1}}$:
\begin{eqnarray}
\alpha = \dfrac{\nu_{ij}}{c} \cdot \sqrt{\dfrac{2\cdot N_A\cdot k_{b}\cdot T\cdot \mathrm{ln}(2)}{M_{mol}}}
\label{GHWHM}.
\end{eqnarray}\\

The Gaussian profile depends mainly on the temperature and not on the pressure. The Gaussian profile is expected to dominate in the upper part of the Earth's atmosphere, where pressure broadening is negligible.

\subsubsection{Lorentzian profile}
The Lorentzian profile is the theoretical profile that describes the pressure broadening of a line: 
\begin{eqnarray}
f_{Lorentzian}(\nu) = \dfrac{1}{\pi}\cdot \dfrac{\gamma}{\gamma^2 + (\nu -\nu_{ij}^{*} )^2}
\label{Lprofile},
\end{eqnarray}
where $\gamma$ is the Lorentzian (pressure-broadened) HWHM that depends on the pressure $P$, the temperature $T$, the partial pressure $P_{self}$, and the reference temperature $T_{ref}$\,=\,296\,K:
\begin{eqnarray}
\gamma = (\dfrac{T_{ref}}{T})^{n_{air}} \cdot (\gamma_{air} \cdot (P_{mean}-P_{self} + \gamma_{self}\cdot P_{self})
\label{LHWHM_full}.
\end{eqnarray}
In this case, $P_{self}$ can be neglected and the HWHM can be simplified to
\begin{eqnarray}
\gamma = \left(\dfrac{296}{T}\right)^{n_{air}} \cdot (\gamma_{air} \cdot P_{mean}).
\label{LHWHM}
\end{eqnarray}

The Lorentzian profile depends mainly on the pressure and not on the temperature. The Lorentzian profile is expected to dominate in the lower part of the Earth's atmosphere, where pressure is higher and, coincidently, where most of the water vapor is located.

\subsubsection{Voigt profile}\label{Vprofile}
The Voigt profile is simply the convolution of the Gaussian and Lorentzian profiles. Formally, it is the profile that should be used to reproduce telluric lines because all lines have a contribution of thermal and pressure broadening:
\begin{eqnarray}
f_{Voigt}(\nu) = f_{Lorentzian}(\nu) \ast f_{Gaussian}(\nu)
\label{voigt}.
\end{eqnarray}

\subsection{HITRAN line list}
\begin{figure}[h]
\resizebox{\hsize}{!}{\includegraphics{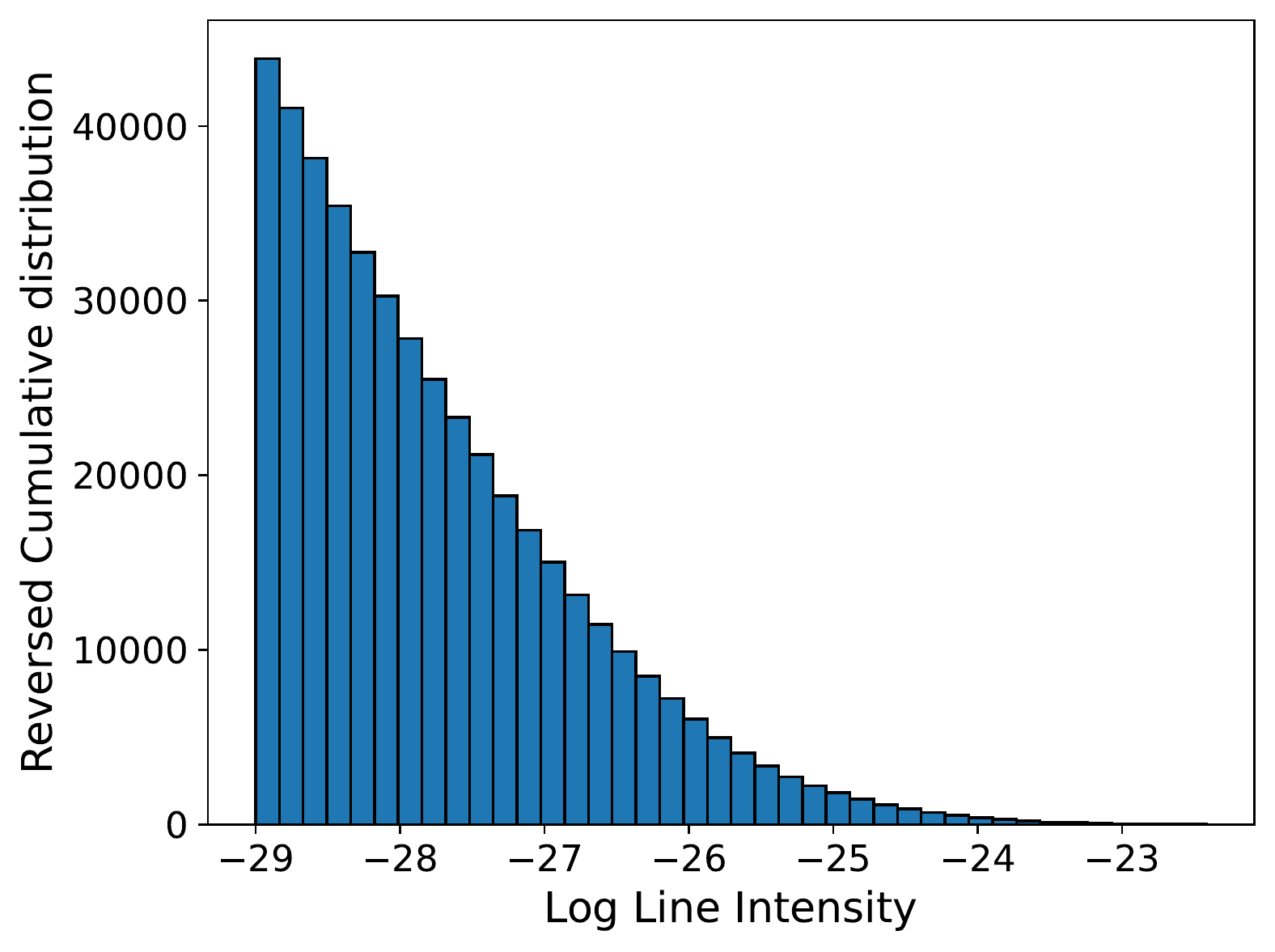}}
\caption{Reversed cumulative distribution of the water vapor line intensities in the water band ranging from 3800 to 8000\,\AA.}
\label{lines_distribution}
\end{figure}

We use the HITRAN \citep{rothman_hitran2012_2013,gordon_hitran2016_2017} database to build the telluric spectrum. Lines from 3800 to 8000 \AA\ for H$_2$\,$^{16}$O that have a line intensity stronger than $10^{-26}$\,$\mathrm{cm^{-1} / (molecule \cdot cm^{-2})}$ are selected. If too few lines are included, the model will not be representative of the observation. When too many lines are included, the weakest lines that do not contribute to the telluric spectrum increase the computation time, as they are the most numerous (see Fig.\,\ref{lines_distribution}). 

\begin{figure}[h]
\resizebox{\hsize}{!}{\includegraphics{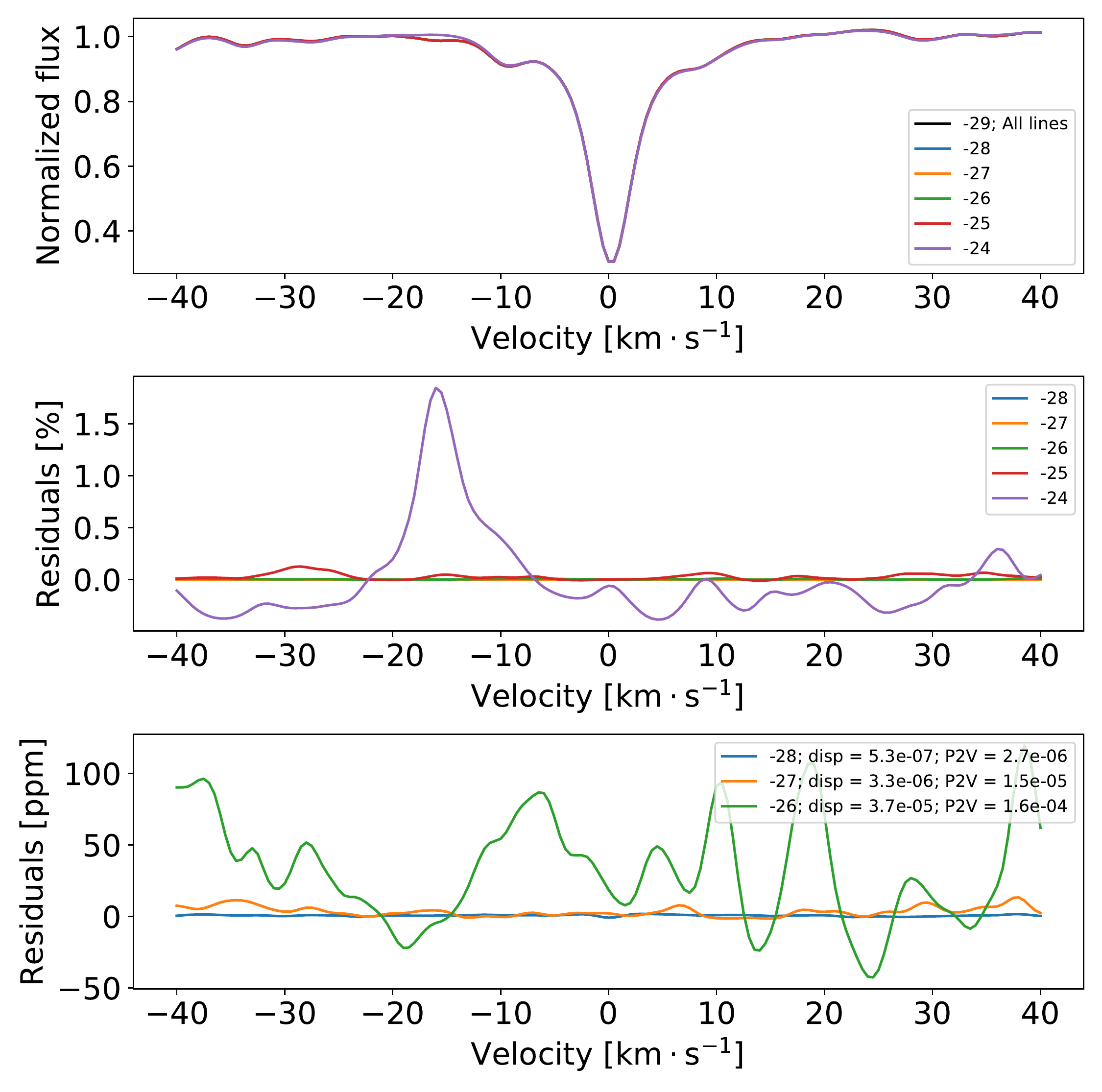}}
\caption{Impact of the number of lines selected in the telluric model. \textit{Top panel:} Average water telluric line built for the model, with line intensity thresholds at $10^{-24}$, $10^{-25}$, $10^{-26}$, $10^{-27}$, $10^{-28}$, and $10^{-29}$\,$\mathrm{cm^{-1} / (molecule \cdot cm^{-2})}$. \textit{Middle panel:} Residuals of the average lines with respect to including all the lines. See Sect.\,\ref{Sec_model_fit} for more details on how the average line is built. \textit{Bottom panel:} Zoom on the residuals obtained with the model that includes  the $10^{-26}$, $10^{-27}$, $10^{-28}$\,$\mathrm{cm^{-1} / (molecule \cdot cm^{-2})}$ strongest water lines. The dispersion and P2V amplitude of the residuals are reported in the legend.}
\label{line_intensity_included}
\end{figure}

The cut-off at $10^{-26}$\,$\mathrm{cm^{-1} / (molecule \cdot cm^{-2})}$ was obtained after including more lines or less lines in the model (Fig.\,\ref{line_intensity_included}), and by looking at the dispersion and peak-to-valley (P2V) amplitude of the residuals. It ensures that we model all telluric lines that would be seen in spectra with S/N$>$10$^5$ per pixel, much higher than typical observations, while keeping computing times reasonable.\\

The following parameters of each line have been extracted: $\nu_{ij}$ (vacuum wavenumber in $\mathrm{cm^{-1}}$), $S_{ij}$ (line intensity in $\mathrm{cm^{-1} / (molecule \cdot cm^{-2})}$), $\gamma_{air}$ (air-broadened HWHM in $\mathrm{cm^{-1} / atm}$), $\gamma_{self}$ (self-broadened HWHM in $\mathrm{cm^{-1} / atm}$), $n_{air}$ (temperature dependence exponent for $\gamma_{air}$  in $\mathrm{cm^{-1}}$), $\delta_{air}$ (air-pressure-induced line shift  in $\mathrm{cm^{-1} / atm}$) and $E"$ (lower state energy in $\mathrm{cm^{-1}}$). Another important parameter from the HITRAN database is the table of the total partition sum value, $Q_T$, for H$_2$\,$^{16}$O, which assigns a $Q_T$ value to each temperature.

\subsection{Sky conditions}
The ESPRESSO FITS header contains the following parameters linked to the target, the local weather station, and the ESO radiometer \citep{kerber_monitoring_2012}: the local temperature $T$ in K, the ground pressure $P$ in hPa (it has to be divided by 1013.2501 to have it in atm), and the integrated water vapor\footnote{The integrated water vapor is also called precipitable water vapor.} $IWV$ measured at the zenith in millimeters, which represents the condensed water column. The $IWV$ in the line of sight can be approximated by
\begin{eqnarray}
IWV_{LOS}= IWV \cdot Airmass
\label{IWVLOS}.
\end{eqnarray}

\subsection{Resolution map}\label{subsection_resolution}
The next step consists in convolving the telluric model with the appropriate instrument line spread function (LSF) to reproduce, in the best possible way, the observed spectrum. The instrumental LSF can be approximated by a Gaussian with its full-width half maximum (FWHM) equal to the instrumental resolution. However, the instrumental resolution is not constant across the spectrum. It depends on the optics used to project the light on the detectors, and on potential mechanical tensions on the detectors. To characterize the instrumental resolution of ESPRESSO across the detectors \citep{pepe_espresso_2021}, the light of a thorium-argon lamp, used for calibration, is injected into the science fiber. Thorium lines are spread across both detectors and are present at several pixel positions in each order. As they are ultra-narrow lines, they are not resolved, and thus their FWHM is a direct measure of the instrumental resolution. Consequently, it is possible to map the detector and derive a resolution map of ESPRESSO for the different observing modes. To obtain a resolution value at each pixel of each order, we applied a two-dimensional polynomial fit, where $x$ corresponds to the pixel of each order $y$:
\begin{eqnarray}
R = a \cdot x + b \cdot y + c \cdot x \cdot y + d \cdot x^2 + e \cdot y^2 + f \cdot x^2 \cdot y^2 + g
\label{eq_resolution}.
\end{eqnarray}

This parametrization was found to best reproduce the instrumental resolution measured on the unresolved Thorium lines. The coefficients can be found in Appendix\,\ref{AppendixA}.\\

Figure\,\ref{MAP_R_HR} shows the resolution map (orders versus pixels) for the high-resolution mode using the 1x1 binning. As described in \cite{pepe_espresso_2021} and due to the high resolution of ESPRESSO combined with the large collecting area of the VLT, a pupil slicer is used to reduce the spectrograph size. Therefore, each physical order is duplicated on the detectors into two slices. The dots are the measured resolution on the unresolved thorium lines, while the background map is the fit applied to those lines. The blue detector corresponds to slices from 0 to 89, while the red detector is for slices 90 to 169. The median resolution is about 138000, but variations are visible between both detectors. The red detector has a higher resolution than the blue one globally but shows some variations toward the bottom left edge. The resolution of the blue detector decreases toward the top left and bottom right edges. Those variations reflect the instrumental LSF but also opto-mechanic adjustments applied to the detectors. It is also important to note that there is some spectral overlap between consecutive orders, and thus the right edge of one order probes the same wavelength as the left part of the order above it. In other terms, one wavelength can be present in two consecutive orders (four slices) with different flux values, but also with different resolutions. 

\begin{figure}[h]
\resizebox{\hsize}{!}{\includegraphics{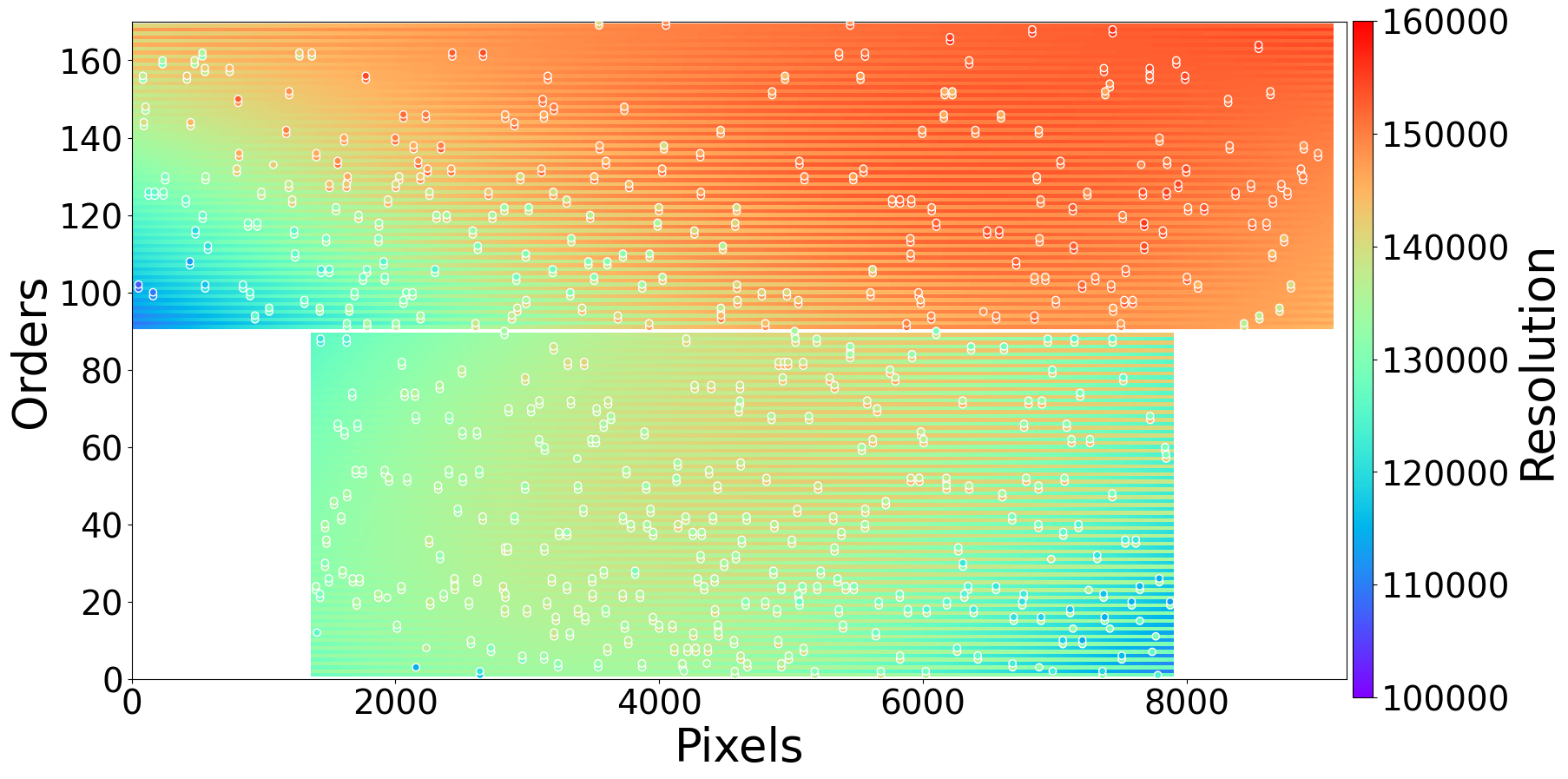}}
\caption{Resolution map (orders versus pixels) for blue and red detectors in the HR1x1 mode. The dots are the measured resolution on individual thorium lines at known positions in pixels and in orders. The background map is the polynomial fit applied to the data.}
\label{MAP_R_HR}
\end{figure}


\section{Telluric model fit to ESPRESSO data}\label{Sec_model_fit}

The algorithm developed to automatically correct the telluric lines consists of adjusting a minimal number of parameters able to describe the spectra in a short computation time\footnote{A couple of minutes on a laptop.}. To do so, the parameters are adjusted on an average telluric line through a CCF, which works as a de facto average line. The use of the CCF has the advantage of reducing stellar contamination, which leads to a more robust determination of the telluric parameters. The produced average line is specific to each molecule and is built over what we refer to hereafter as the subset of telluric lines. The subset is built such that it consists of the strongest unsaturated lines in a defined wavelength range and such that no lines are closer than 80\,$\mathrm{km \cdot s^{-1}}$ to each other. In the case of H$_2$O in the spectral range of ESPRESSO, we consider the subset to be composed of the 20 strongest lines (see Appendix\,\ref{AppendixB}) after excluding the lines at 7186.51, 7208.41, and 7236.72\,\AA, which are removed as they can be saturated depending on the atmospheric water content.  \\

The algorithm starts by identifying the orders or slices (with a spectral coverage reduced by 160\,$\mathrm{km \cdot s^{-1}}$ on each edge\footnote{This is done to ensure the complete modeling of the telluric lines and to avoid numerical issues during the computation of the CCFs.}) where the subset of telluric lines falls. In the case of water vapor for ESPRESSO, the orders are 77 and 78 (slices 154, 155, 156, and 157). For each order or slice, the HITRAN parameters are extracted for the lines of the subset present in the wavelength range and for all lines in a passband of 80\,$\mathrm{km \cdot s^{-1}}$ around them. Then, the telluric spectrum is computed as in Sect.\,\ref{Sec_model_description}. The next step consists in computing the CCF with a telluric binary mask covering the wavelength range. The mask is composed of the subset of telluric lines, for which their weight is fixed to one. The CCF is computed over both the observed stellar spectrum and the telluric model, afterward called observed and modeled telluric CCFs. These observed and modeled telluric CCFs are co-added over the selected orders or slices to obtain a master observed telluric CCF and a master modeled telluric CCF. Both CCFs are then normalized by their continuum. A Levenberg-Marquardt algorithm is then applied to minimize the residuals between the normalized master observed telluric CCF and the normalized master modeled telluric CCF (the free parameters of the model are discussed below). Finally, the best-fit parameters are used to compute the telluric model over the full spectral range of the observation, as in Sect.\,\ref{Sec_model_description}.

\subsection{Impact of line profile}
It is essential to assess the impact of the line profile and to select the one that best models the data with the fewest parameters. The fitted parameters are $T$ and $IWV_{LOS}$ for the Gaussian profile (Eq.\,\ref{Gprofile}), and $T$, $P_{mean}$, and $IWV_{LOS}$ for both the Lorentzian (Eq.\,\ref{Lprofile}) and Voigt (Section\,\ref{Vprofile}) profiles.

\begin{figure}[h]
\resizebox{\hsize}{!}{\includegraphics{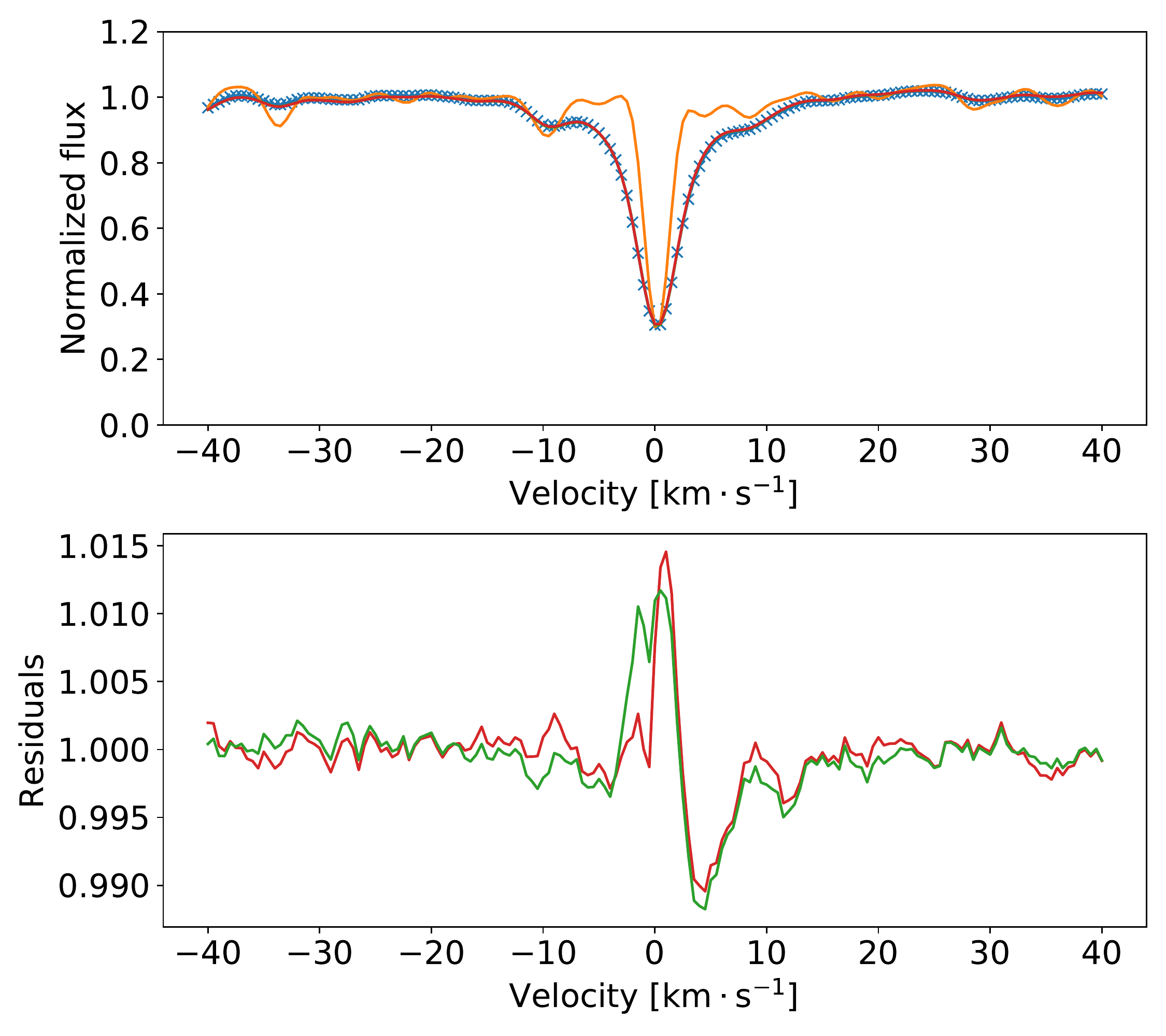}}
\caption{Impact of the line profile to best model the data. \textit{Top panel:} CCFs computed over the 20 strongest water lines. The blue CCF is the master observed telluric CCF based on ESPRESSO data, while the orange, green, and red CCFs are the master modeled telluric CCFs assuming Gaussian, Lorentzian, or Voigt profiles, respectively. \textit{Bottom panel:}  Residuals only for the Lorentzian and Voigt profiles. The Gaussian profile is not shown due to its larger residuals.}
\label{line_profile}
\end{figure}

Figure\,\ref{line_profile} compares the normalized master observed telluric CCF obtained with the subset of telluric lines (in blue) to the master modeled telluric CCFs based on the three different profiles studied. The Gaussian profile is not able to reproduce the observed average water telluric line because it does not account for the pressure broadening. The two other models reproduce the data well, with residuals below the 2\% P2V amplitude (the residual shape is discussed in Sect.\,\ref{Sect_HR1544}). This threshold of  2\,\% P2V will be used to assess the data correction quality in the following tests. This choice represents the threshold for micro-telluric lines that are not historically masked by the DRS of HARPS or ESPRESSO. In the case of water vapor, the contribution of the Gaussian profile in the Voigt profile is negligible, and to follow Occam's razor principle, the Lorentzian profile was selected as it is the simplest one.\\

\subsection{Impact of the temperature}
\begin{figure}[h]
\resizebox{\hsize}{!}{\includegraphics{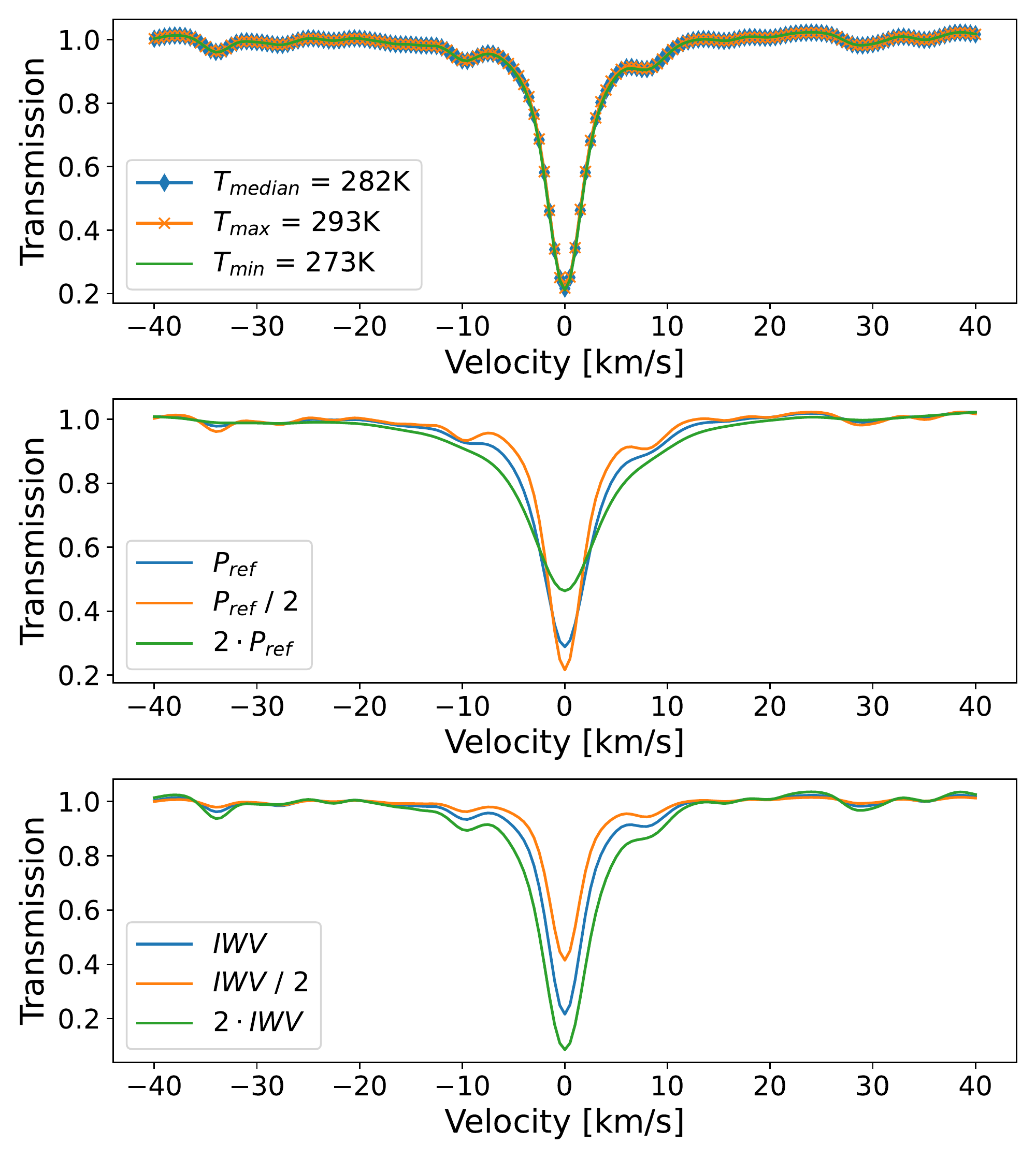}}
\caption{Impact of $T$ (\textit{top}), $P$ (\textit{middle}), and $IWV_{LOS}$ (\textit{bottom}) on the telluric line shape. \textit{Top:} Master modeled telluric CCF at an atmospheric temperature of 282\,K (blue), 293\,K (orange) and 273\,K (green). $P_{mean}$ and  $IWV$ are fixed to the same values for the three models. \textit{Middle:} Master modeled telluric CCF at ground pressure (blue), half of the ground pressure (orange), and twice the ground pressure (green). $T$ and  $IWV_{LOS}$ are fixed to the same values for the three models. \textit{Bottom:} Master modeled telluric CCF at $IWV_{LOS}$ (blue), half of $IWV_{LOS}$  (orange), and twice $IWV_{LOS}$ (green). $T$ and $P_{mean}$ are fixed to the same values for the three models.}
\label{properties_variation}
\end{figure}

Figure\,\ref{properties_variation} (top panel) shows the impact of the temperature on the Lorentzian line shape, while $P_{mean}$ and $IWV$ are respectively fixed to half the ground pressure and $IWV_{LOS}$. The temperature was set at the lowest (273\,K), the median (282\,K), and the highest (293\,K) temperatures registered since the start of ESPRESSO operations. The temperature does not have any impact on the line shape and it can be fixed to the ground temperature measured at the time of the observations.


\subsection{Impact of the pressure}
Figure\,\ref{properties_variation} (middle panel) shows the impact of the pressure on the Lorentzian line shape, while $T$ and $IWV$ are respectively fixed to the ground temperature and $IWV_{LOS}$. The pressure was set to the ground pressure, half of the ground pressure, and twice the ground pressure. When the pressure decreases, the line shape is narrower and deeper. Contrary to the temperature, the pressure plays an important role in the line shape and the algorithm fits it. 


\subsection{Impact of the integrated water vapor}
Figure\,\ref{properties_variation} (bottom panel) shows the impact of the $IWV$ on the Lorentzian line shape, while $T$ and $P_{mean}$ are respectively fixed to the ground temperature and half the ground pressure. The \textit{IWV} was scaled according to airmass, $IWV_{LOS}$, half of $IWV_{LOS}$, and twice $IWV_{LOS}$. When the water content increases, the line is deeper. The \textit{IWV} affects the line as a scaling factor, and it is thus necessary to fit it.



\section{Application of the telluric correction for different spectral types}\label{Sec_App_Stellar_type}
We first performed a test on the spectrophotometric standard star, HR\,1544. As it is an A-type star, contamination by stellar lines is negligible as only a few broad lines are present, while telluric lines are narrower. We then tested the telluric correction for other spectral types up to late-type stars to see if the presence of stellar lines biases the telluric correction.

\subsection{HR\,1544}\label{Sect_HR1544}
\begin{figure*}[t]
\includegraphics[width=\textwidth]{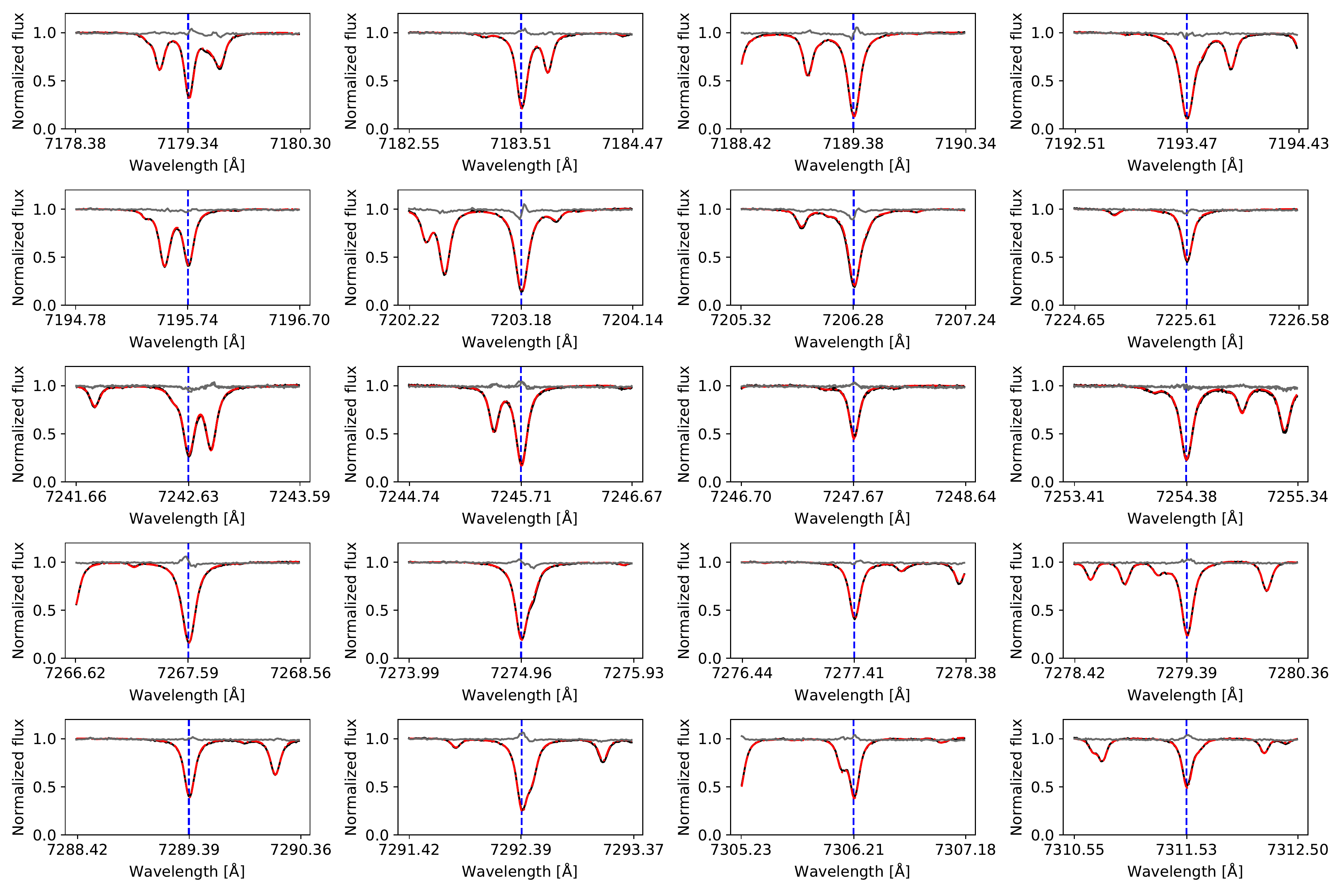}
\caption{Stellar spectrum of HR\,1544 before (black) and after (gray) telluric correction around the subset of the 20 selected lines. Each panel is for a different line, with its position indicated by the vertical dashed blue line. The telluric spectrum is shown as the dashed red curve.}
\label{HR1544_lines_fit_H2O}
\end{figure*}

\begin{figure*}[h]
\includegraphics[width=\textwidth]{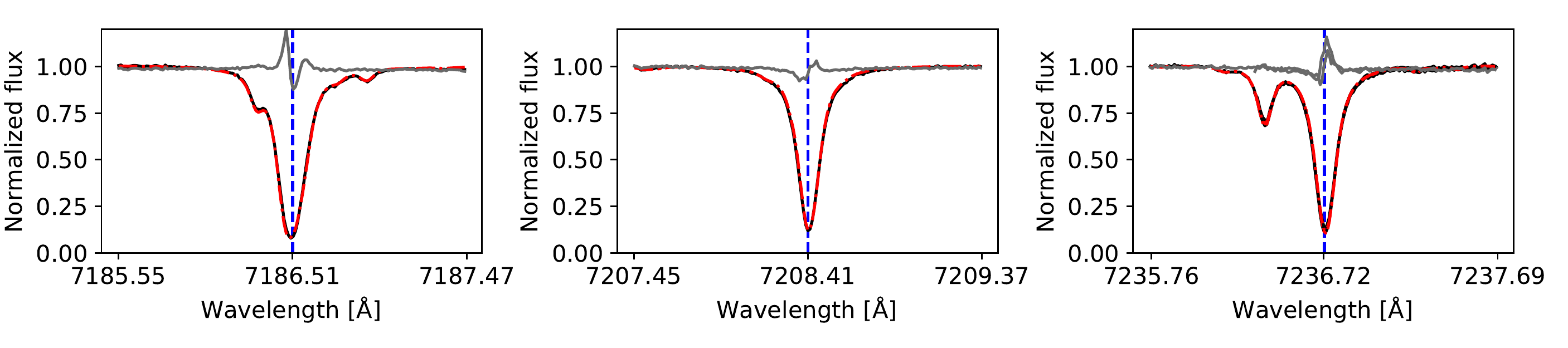}
\caption{Same legend as for Fig.\ref{HR1544_lines_fit_H2O} but for the three lines excluded from the subset.}
\label{HR1544_lines_excluded_H2O}
\end{figure*}

HR\,1544 is an A-type star and is part of the spectrophotometric standard catalog of ESO for ESPRESSO. We selected a spectrum at a high S/N ($\sim$350) with an integrated water vapor of 3.56\,mm measured at the zenith by one of the radiometers installed on Paranal \citep{kerber_monitoring_2012}, which is above the median value of 2\,mm\footnote{https://www.eso.org/sci/facilities/paranal/astroclimate/paranal-figs.html}. Figure\,\ref{HR1544_lines_fit_H2O} shows this spectrum around the subset of lines selected to perform the optimization of the parameters before and after applying the telluric correction, as well as the computed telluric spectrum. The selected lines acquired under this weather condition have an absorption depth ranging from 50 to 80\%. Some systematic residuals are present after the telluric correction around the deepest lines but with a decreasing residual amplitude for the weakest telluric lines. The lines shown are the strongest in the ESPRESSO spectral band, except for the lines at 7186.51, 7208.41, and 7236.72 \AA. Figure\,\ref{HR1544_lines_excluded_H2O} displays them alongside their strong residuals. We note that their residuals have different structures from one line to another and that it was not possible to properly model them, leading to their exclusion. The lack of proper modeling due to (quasi-)saturation can be explained by several factors. First, the one-layer model described here is too simple for strong telluric lines. For example, asymmetric broadening could introduce P-Cygni-like profiles. Secondly, some physics of the lines is missing to reproduce their core. It could come from the inclusion of wind that shifts the line core, or some HITRAN parameters that are inaccurate. Finally, it could be a systematic error caused by the noncommutativity of multiplication and convolution, which can be significant for deep telluric lines.

\begin{figure*}[h]
\resizebox{\hsize}{!}{\includegraphics{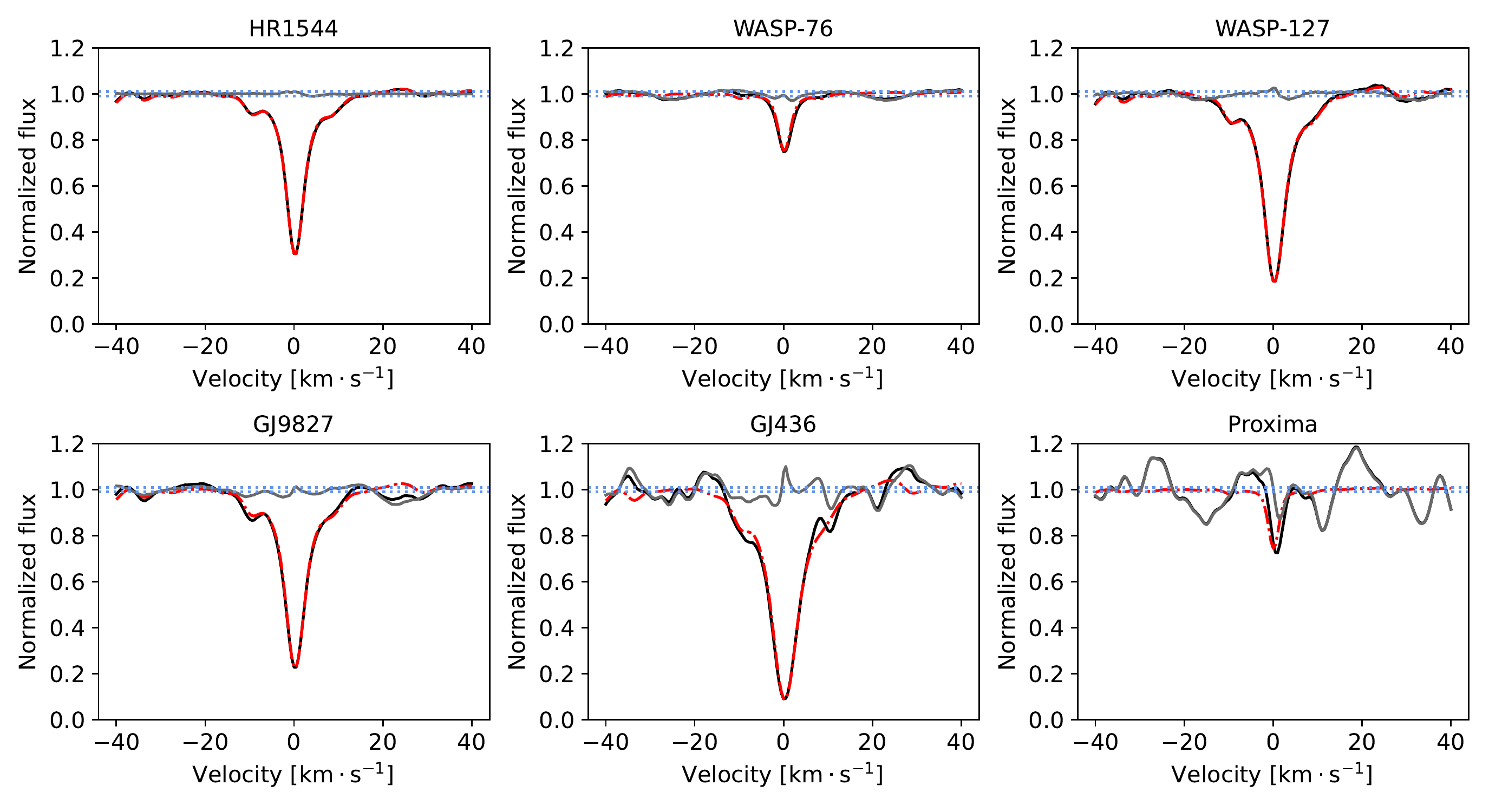}}
\caption{Master observed telluric CCF (in black), the master modeled telluric CCF (in red, dashed) and the master observed telluric-corrected  CCF (in gray) for HR\,1544, WASP-76, WASP-127, GJ\,9827, GJ\,436, and Proxima. The CCFs are performed over the subset of selected water lines. The two horizontal blue dotted lines encompass the 2\% residual P2V.}
\label{ccf_H2O}
\end{figure*}

We used the subset of selected lines shown in Fig.\,\ref{HR1544_lines_fit_H2O} to create the master observed telluric CCF, the master modeled telluric CCF, and the master observed telluric-corrected CCF shown in Fig\,\ref{ccf_H2O}. They represent the average line profile in the 80\,$\mathrm{km \cdot s^{-1}}$ surrounding each of the selected lines. One can see that the Levenberg-Marquardt minimization converges to residuals below the 2\% P2V amplitude on the average line. The latest value is in agreement with typical telluric correction from the literature \citep[e.g.,][]{sameshima_correction_2018,ulmer-moll_telluric_2019,baker_iag_2020}.

\subsection{From F-type stars to K-type stars}

Figure\,\ref{ccf_H2O} also shows the observed telluric CCF, the master modeled telluric CCF, and the master observed telluric-corrected CCF for WASP-76, WASP-127, and GJ\,9827, which are F-, G-, and K-type stars, respectively. The three stars are part of the observations of the ESPRESSO consortium with an individual S/N of  $\sim$130 and with an IWV of 0.7, 7, and 5\,mm respectively. It is noticeable that the IWV not only plays an important role in the absorption line, as expected, but it also leads to stronger residuals in the very core of telluric lines. Nonetheless, the residuals are within or close to the 2\% P2V amplitude and in agreement with the continuum dispersion. The later type stars have more and stronger stellar lines that slightly affect the CCF continuum, such as at $\sim$30\,$\mathrm{km \cdot s^{-1}}$ for WASP-127 or $\sim$25\,$\mathrm{km \cdot s^{-1}}$ for GJ\,9827, but do not prevent the recovery of the water line profile and the correction of telluric contamination.

\subsection{The case of M-type stars}
M-type stars are rich in molecular bands in their visible spectrum, causing absorption of the stellar continuum to an extent that what we measure is de facto a pseudo-continuum. It is thus challenging to disentangle stellar and telluric lines. Figure\,\ref{ccf_H2O} also includes the observed telluric CCF, the master modeled telluric CCF, and the master observed telluric-corrected CCF for GJ\,436 and Proxima, which are M2 and M5-type stars, respectively. It is not possible to discriminate between a telluric residual and the dispersion of the pseudo-continuum in the master observed telluric-corrected CCF. The spectrum of GJ\,436 was obtained in extreme conditions with an IWV of 12\,mm, leading to an absorption depth of about 90\%, and thus a low S/N in the telluric line cores. On the contrary, the spectrum of Proxima was obtained under dry conditions with an IWV of 0.3\,mm, leading to a telluric correction at the level of the pseudo-continuum dispersion.\\

From these examples, it is possible to conclude that the telluric model presented here can be applied automatically to all observations of ESPRESSO\footnote{Tests on QSO spectra have to be performed but we expect the correction to perform equally good.}, and it is not biased by the stellar spectrum and contamination between stellar and telluric lines. Nonetheless, some telluric residuals can be present in the core of the strongest lines but over a very narrow spectral range in comparison to the whole spectral domain impacted by water vapor telluric lines.

\section{Application of the telluric correction for radial velocity}\label{Sec_App_RV}

To assess the impact of the telluric correction on the RVs obtained with ESPRESSO, we corrected the spectra of Tau Ceti and Proxima from telluric contamination by dividing each spectrum by its best-fit telluric template. Then, the spectra were cross-correlated with the DRS masks as the data without telluric correction. The next step was to see if a broader spectral coverage could be used. \\
It is essential to understand how ESPRESSO stellar masks are built. Each mask corresponds to a given spectral type (or sub spectral type for M dwarfs) but they are not specific to a given star. The masks are built from high S/N spectra of inactive stars observed with ESPRESSO. Firstly, the line locations are identified by searching for local minima. Then, spectral regions affected by magnetic activity stellar lines (such as \ion{Ca}{II}\,H $\&$\,K, Na, and H-$\alpha$) and by strong telluric bands (5871-6005\,\AA, 6271-6341\,\AA, 6449-6605\,\AA, 6861-7416\,\AA, 7588-7751\,\AA, and 7860-8000\,\AA) are excluded. As the masks need to be the same, independently of the star or epoch of the year, margins of 64\,km$\cdot$s$^{-1}$ (1.5\,\AA\ at 7010\,\AA) for active stellar lines and 214\,km$\cdot$s$^{-1}$  (5\,\AA\ at 7010\,\AA) for telluric bands, to include systemic velocity and barycentric shifts, are excluded.\\
Therefore, to understand how the telluric correction could help to reduce these excluded regions, we produced new masks based on telluric-free spectra for Tau Ceti and Proxima. Then, we modified the exclusion regions as defined by the mask in different steps to understand how the new spectral regions modify the RV time series. The different RV time series produced for each star are named \texttt{DRS}, \texttt{Microtell}, \texttt{Blue}, \texttt{Orders77\_78}, \texttt{Orders77}, \texttt{15\_Slines}, \texttt{1\_Sline}, and \texttt{Ideal}. The associated masks aer shown in Figs.\,\ref{mask_applied} and\,\ref{mask_applied_zoom}.

\texttt{DRS} is the standard RV time serie with no telluric correction and no modification of the mask. It is important to note that this is not necessarily the best mask as, for example, micro-telluric lines can have an impact on the shape of the CCF. \\
\texttt{Microtell} is the RV time serie obtained with the data corrected for the telluric lines but using the same mask as \texttt{DRS}. These data show the impact of micro-telluric lines on the shape of the CCFs, and therefore on the extracted RVs.\\
\texttt{Blue} is the RV time serie obtained with the data corrected for the telluric lines but the mask includes the wavelength regions with the telluric water bands around the Na and H-$\alpha$ lines, and the reddest water band.\\
\texttt{Orders77\_78} is the RV time serie obtained with the data corrected for the telluric lines but the mask includes the wavelength regions with all the telluric water bands. It is important to note, however,  that both orders  77 and 78 (slices 154, 155, 156, and 157), where water telluric lines are the strongest, and regions containing O$_2$ lines are still excluded, as in the standard mask. \\
\texttt{Orders77} is the RV time serie obtained with the data corrected for the telluric lines and the mask includes the wavelength regions with all the telluric water bands, but the physical order 77 (slices 154, 155) is still excluded in addition to the oxygen bands.\\
\texttt{15\_Slines} is the RV time serie obtained with the data corrected for the telluric lines and the mask includes the wavelength regions with all the telluric water bands except the 15 strongest water lines with an exclusion window of 214\,km$\cdot$s$^{-1}$. \\
\texttt{1\_Sline} is the RV time serie obtained with the data corrected for the telluric lines and the mask includes the wavelength regions with all the telluric water bands except the strongest water line at 7186.51\,\AA\ with an exclusion window of 214\,km$\cdot$s$^{-1}$ around each line.\\
\texttt{Ideal} is the RV time serie obtained with the data corrected for the telluric lines and the mask includes the wavelength regions with all the telluric water bands. It corresponds to the mask that should be used if there were no telluric water lines or if their correction was assumed to be perfect (i.e., affecting the photon noise without introducing systematic in the RVs).\\

We also note that ESPRESSO had several technical interventions introducing possible offsets at 2458662 and 2459202 BJD \citep{faria_candidate_2022}.
\subsection{Tau Ceti}
Tau Ceti (V-mag=3.5) is one of the most studied stars and has been heavily observed by high-precision RV campaigns with HARPS and now ESPRESSO \citep[e.g.,][]{dumusque_planetary_2011,feng_color_2017}. Between July 2018 and December 2022, 1982 data points across 202 nights were acquired with ESPRESSO. The observational strategy for Tau Ceti consists in taking several very short exposures per night. Therefore, the results presented below are nightly binned. Tau Ceti RVs are known to be highly stable with no clear sign of activity nor any massive planets (more massive than super-Earths). It is thus a perfect target for studying the impact of the telluric correction. \\

\begin{figure}[h]
\resizebox{\hsize}{!}{\includegraphics{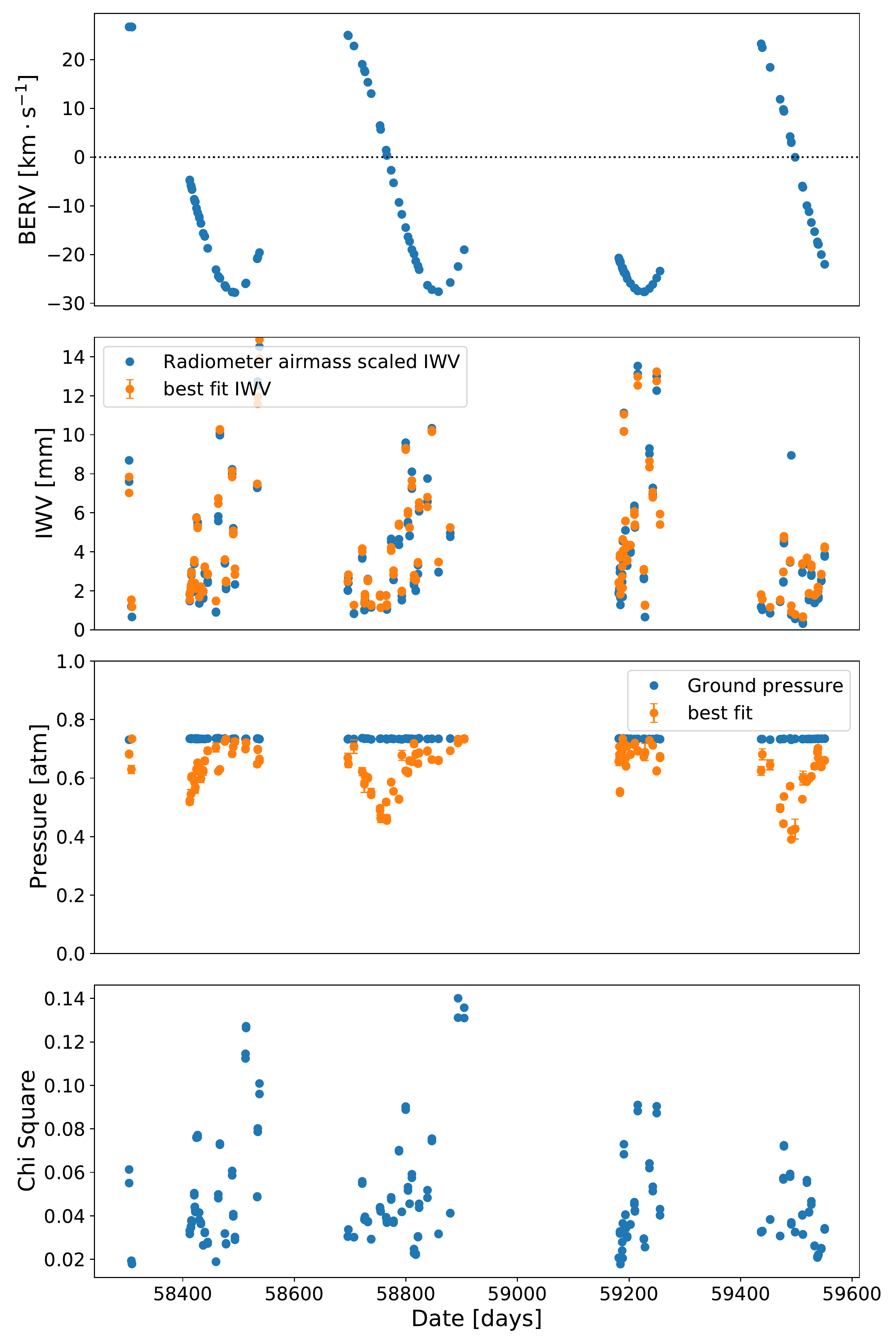}}
\caption{Properties of the telluric correction best-fit. \textit{First panel:} Barycentric correction as a function of time for Tau Ceti observations. \textit{Second panel:} $IWV$ as a function of time. In blue is the airmass scaled $IWV_{LOS}$ as measured by the radiometer and in orange is the best-fit IWV value. \textit{Third-panel:} Pressure as a function of time. In blue is the measured ground pressure (it is the upper limit to the fitted pressure) and in orange is the best-fit pressure value. \textit{Last panel:} Best chi-square values of the fit of the automatic telluric correction.}
\label{TauCeti_tell}
\end{figure}

\begin{figure}[h]
\includegraphics[width=0.5\textwidth]{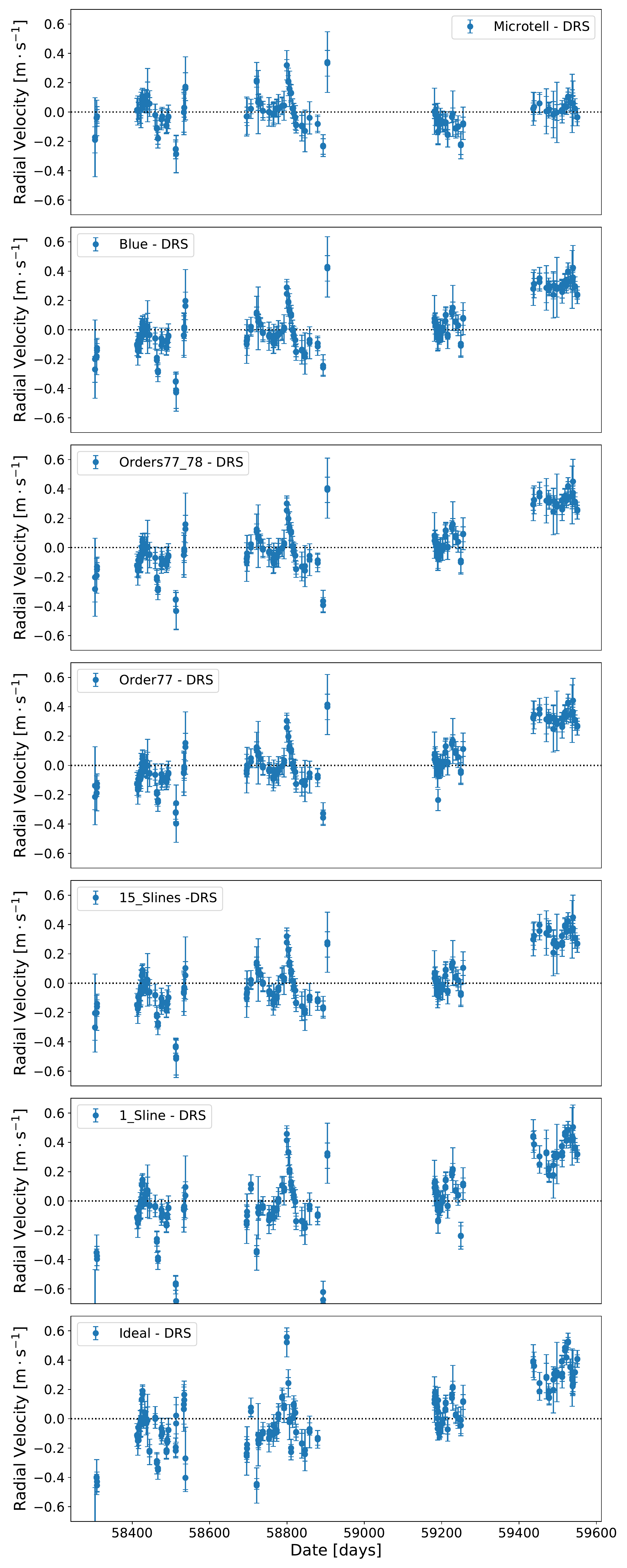}
\caption{Tau Ceti RV time series differences between \texttt{DRS} and, from top to bottom: \texttt{Microtell}, \texttt{Blue}, \texttt{Orders77\_78}, \texttt{Order77}, \texttt{15\_Slines}, \texttt{1\_Sline}, and \texttt{Ideal}.}
\label{TauCeti_comp}
\end{figure}

\begin{figure}[h]
\includegraphics[width=0.5\textwidth]{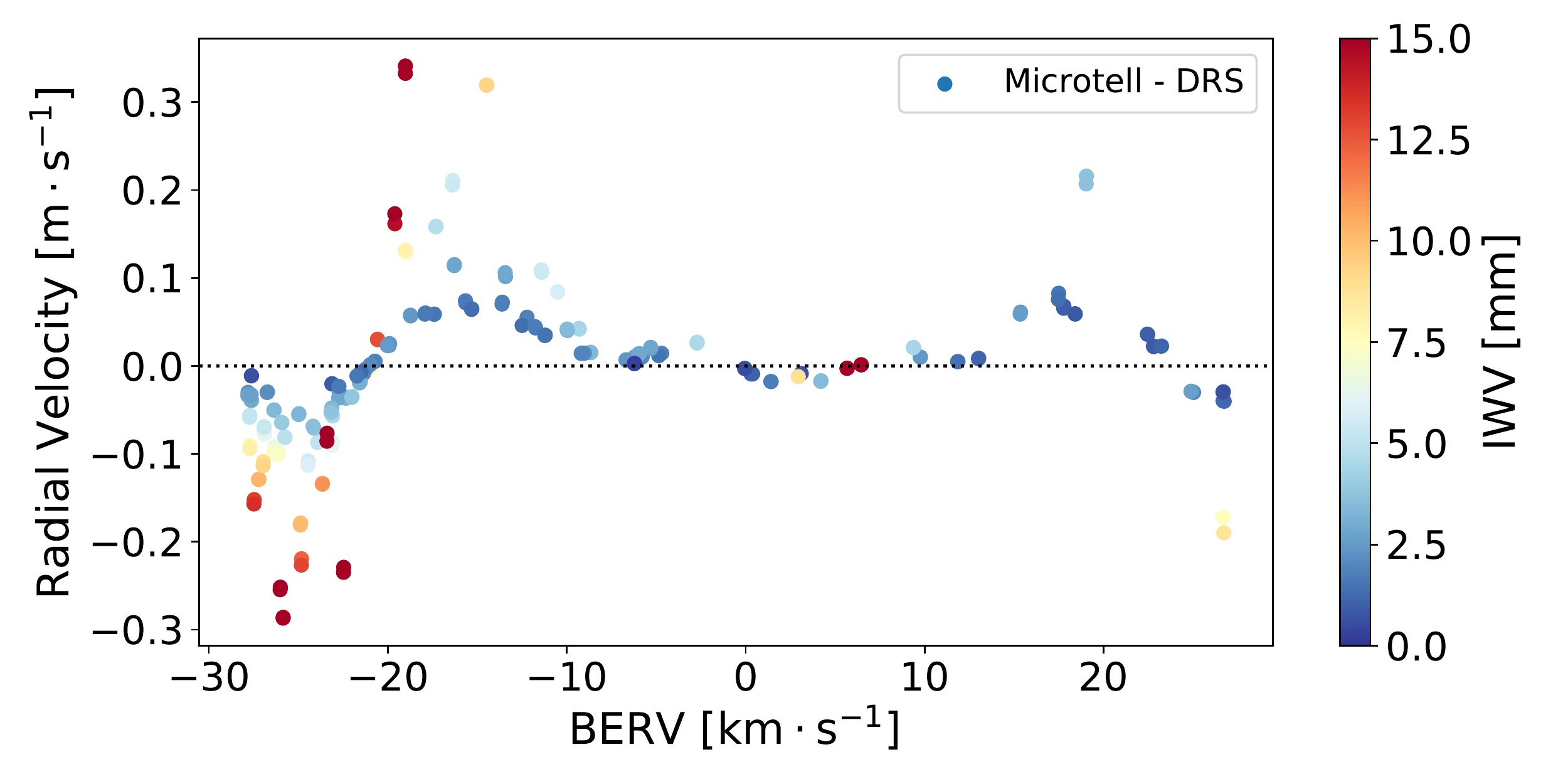}
\caption{Tau Ceti RV time series differences between \texttt{DRS} and \texttt{Microtell} as a function of BERV and colored by the $IWV$ content.}
\label{TauCeti_phase}
\end{figure}

Figure\,\ref{TauCeti_tell} shows the best-fit parameters of $IWV$ and pressure with the corresponding chi-square for Tau Ceti. Our computed $IWV$ values are in agreement with the radiometer values, and the slight differences (rms of the difference is $\sim$0.4\,mm) can be explained by different water column distributions between the line of sight and the zenith. The derived pressure varies from half of the ground pressure up to the ground pressure\footnote{The ground pressure is the upper limit to the fitted pressure.}, with seasonal variations. Finally, we note that the quality of the fits is stable over time, confirming the capability of our technique to be applied automatically. Figure\,\ref{TauCeti_comp} shows the difference between the RV time series obtained with the different masks after the telluric correction is applied, and the RV time series obtained with the \texttt{DRS}. One can see that the \texttt{Microtell} RV time series has some systematic differences in comparison to the \texttt{DRS} RV time series. These variations are due to the presence of micro-telluric lines that induce RV variations of about 58\,cm$\cdot$s$^{-1}$ from P2V with an rms of 10\,cm$\cdot$s$^{-1}$. When phased as a function of the BERV, the residuals also show a periodic contribution (Fig.\,\ref{TauCeti_phase}) as expected by the Earth's orbital motion around the Sun. The same structures are also present with the other masks. We confirmed the micro-telluric contribution by computing the RVs for the \texttt{DRS} and \texttt{Microtell} RV time series in the spectral range from 4950-5855\,\AA, where the micro-telluric lines are located (see Fig.\,\ref{ESPRESSO_tell}). We observed the exact same systematics but with an amplified signal up to 1.4\,m$\cdot$s$^{-1}$. It is thus proof that Tau Ceti spectra are impacted by micro-telluric contamination. Moreover, the \texttt{Order77}, \texttt{15\_Slines}, \texttt{1\_Sline}, and \texttt{Ideal} RV time series have increasing systematics when a broader spectral range is included. They are probably due to telluric residuals around the 7250\,\AA\ water band. The large difference between the \texttt{1\_Sline} and \texttt{Ideal} RV time series is  caused by a single telluric line residual (at 7186.51\,\AA). This shows how important it is to exclude telluric lines or to be very confident in the level of correction achievable. We also note an RV offset for the last epoch of observations after 2459202\,BJD, which coincides with a technical intervention as mentioned earlier. To conclude, the mask \texttt{Orders77\_78} is selected for G-type stars, such as Tau Ceti, as it is a good trade-off between reducing bias from telluric residuals and the gain in photon noise ($\sim$2\,\%, Table\,\ref{TauCeti}). The slight gain in photon noise is caused by the blackbody maximum position at blue wavelengths for G-type stars, where few spectral regions are excluded.

\begin{table}[h]
\centering
\caption{Summary of Tau Ceti masks and photon noise.}
\begin{tabular}{lccc}
\hline
      Mask      & \makecell{Spectral range \\ covered [\AA]} & \makecell{Number \\ of lines} & \makecell{Photon noise \\ precision [cm$\cdot$s$^{-1}$]} \\ 
\hline
\texttt{DRS} &        2948        &      5646       & 4.39 \\ 
\texttt{Microtell} &        2948        &      5646       & 4.38 \\ 
\texttt{Blue}      &        3255        &      5811       & 4.31\\ 
\texttt{Orders77\_78}   &        3477        &      5943       & 4.29 \\ 
\texttt{Order77}    &        3562        &      5993       & 4.28 \\ 
\texttt{15\_Slines}    &        3576        &      6006       & 4.28 \\ 
\texttt{1\_Sline}     &        3672        &      6058       & 4.26 \\ 
\texttt{Ideal}      &        3682        &      6069       & 4.26 \\ 
\hline
\end{tabular}
\label{TauCeti}
\end{table}

\subsection{Proxima}
Proxima is also one of the most studied stars and has been heavily observed by high-precision RV campaigns with HARPS and ESPRESSO. Proxima is the closest star to the Sun and is orbited by an Earth-like planet at a period of 11\,days \citep{anglada-escude_terrestrial_2016}, and by a planet candidate at a large orbital distance \citep{damasso_low-mass_2020}. Between February 2019 and August 2021, 117 data points were acquired and led to the independent confirmation of Proxima b by ESPRESSO \citep{suarez_mascareno_revisiting_2020}, and the detection of Proxima d at an orbital period of 5.12 days \citep{faria_candidate_2022}. As the star is an M5 dwarf, its spectral energy distribution peaks toward red wavelengths, where the strongest telluric lines are located. Therefore, it is a perfect target for verifying if the telluric correction can lead to improved precision by including more stellar lines in the CCFs.\\

\begin{figure}[h]
\resizebox{\hsize}{!}{\includegraphics{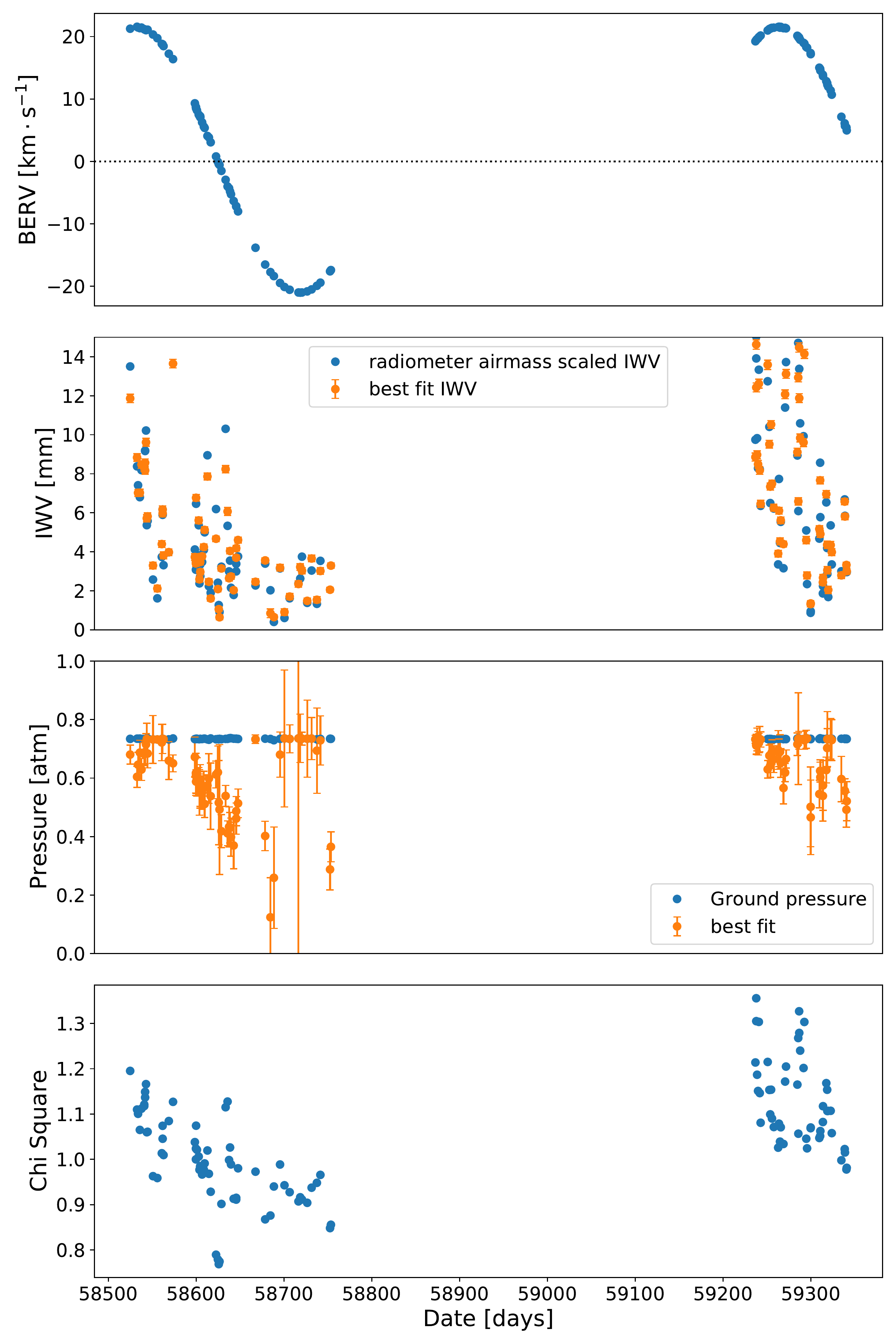}}
\caption{Properties of the telluric correction best-fit. \textit{First panel:} Barycentric correction as a function of time for Proxima observations. \textit{Second panel:} $IWV$ as a function of time. In blue is the airmass scaled $IWV_{LOS}$ as measured by the radiometer and in orange is the best-fit IWV value. \textit{Third panel :} Pressure as a function of time. In blue is the ground pressure (it is the upper limit to the fitted pressure) and in orange is the best-fit pressure value. \textit{Last panel:} Best chi-square values of the fit of the automatic telluric correction.}
\label{Proxima_tell}
\end{figure}

\begin{figure}[h]
\includegraphics[width=0.5\textwidth]{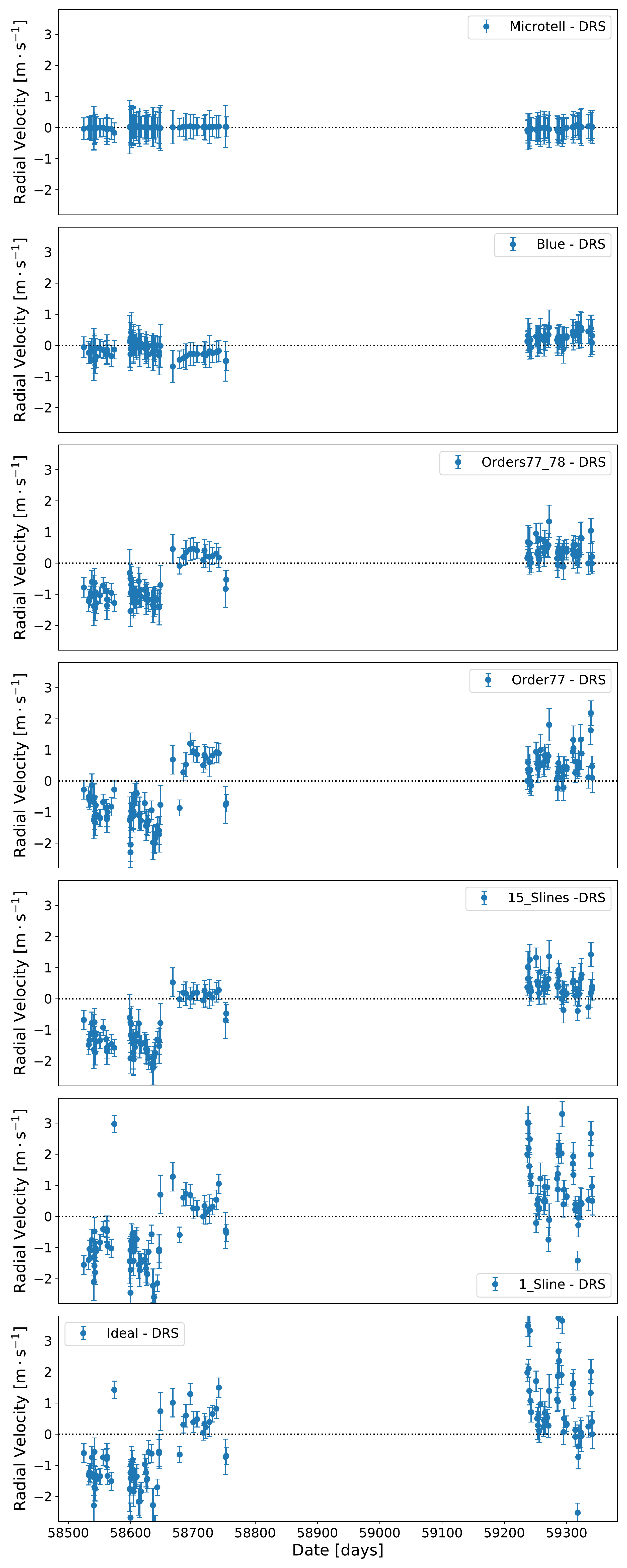}
\caption{Proxima RV time series differences between \texttt{DRS} and, from top to bottom: \texttt{Microtell}, \texttt{Blue}, \texttt{Orders77\_78}, \texttt{Order77}, \texttt{15\_Slines}, \texttt{1\_Sline}, and \texttt{Ideal}.}
\label{Proxima_comp}
\end{figure}

Figure\,\ref{Proxima_tell} shows the best-fit parameters of $IWV$ and pressure with the corresponding chi-square for Proxima. We confirm the same behaviors for the $IWV$ (rms of the difference is $\sim$0.7\,mm), the pressure, and the fit goodness as for Tau Ceti. Figure\,\ref{Proxima_comp} shows the difference between the RV time series obtained with the different masks after the telluric correction is applied, and the RV time series obtained with the \texttt{DRS}. One can see that the \texttt{Microtell} and \texttt{Blue} RV time series are similar to the \texttt{DRS} RV time series. This is expected as the micro-telluric lines and the spectral range gained with the \texttt{Blue} mask are in the blue part of the ESPRESSO range, where there is very little flux for a star such as Proxima. The \texttt{Orders77\_78} RV time series differs from the \texttt{DRS} RV time series by an offset for the RVs obtained after 2458650 BJD but no systematics are introduced. This RV offset coincides with a technical intervention on ESPRESSO. On the contrary, \texttt{Order77}, \texttt{15\_Slines}, \texttt{1\_Sline,} and \texttt{Ideal} RV time series have clear systematics compared to the \texttt{DRS} RV time series, which increase when a greater spectral range is included. Even if the \texttt{DRS} time series is not the reference, introducing systematics up to 4\,m$\cdot$s$^{-1}$ is not acceptable. Therefore, the telluric correction is not good enough in the core of the 7200\,\AA\ water band for it to be used for precise RV studies. Nonetheless, from Table\,\ref{Proxima}, the gain in photon noise between the time series \texttt{DRS} and \texttt{Orders77\_78} is about 25\,\%, while the gain between the time series \texttt{Orders77\_78} and \texttt{Ideal} is negligible. This factor of 1.33 in photon noise between 28 and 21\,cm$\cdot$s$^{-1}$ corresponds to a factor of 1.78 in exposure time. In other words, using the \texttt{Orders77\_78} RVs is equivalent to increasing the exposure time of the \texttt{DRS} RVs by 1.78 to reach a photon noise of 21\,cm$\cdot$s$^{-1}$. On the contrary, reaching the same photon noise as the \texttt{DRS} RVs (28\,cm$\cdot$s$^{-1}$) is equivalent to a gain of 44\,\% in exposure time on the \texttt{Orders77\_78} RVs. To conclude on the mask selection, the best mask to use for Proxima is \texttt{Orders77\_78}.

\begin{table}[h]
\caption{Summary of Proxima masks and photon noise.}
\begin{tabular}{lccc}
\hline
      Mask      & \makecell{Spectral range \\ covered [\AA]} & \makecell{Number \\ of lines} & \makecell{Photon noise \\ precision [cm$\cdot$s$^{-1}$]} \\ 
\hline
\texttt{DRS}  &        2948        &     8037       & 28  \\ 
\texttt{Microtell} &        2948        &     8037       & 28  \\ 
\texttt{Blue}      &        3255        &      8978       & 25  \\ 
\texttt{Orders77\_78}   &        3477        &      9666       & 21 \\ 
\texttt{Order77}    &        3562        &      9912       & 20  \\ 
\texttt{15\_Slines}    &        3576        &      9969       & 20  \\ 
\texttt{1\_Sline}     &        3672        &     10246       & 19  \\ 
\texttt{Ideal}      &        3682        &       10276      & 19  \\ 
\hline
\end{tabular}
\label{Proxima}
\end{table}

\subsection{Impact on the parameters of Proxima b and d}
To better understand the impact of the telluric correction on the RV delivered by the ESPRESSO DRS, we measured the impact of the telluric correction on the CCFs for the recovered orbital parameters of Proxima b and d, considering the same model used in \cite{faria_candidate_2022}. We compared the standard RVs produced by the DRS (hereafter referred to as DRS RVs) to the RVs computed with the \texttt{Orders77\_78} mask on the telluric-corrected spectra (hereafter referred to as telluric-corrected RVs). This analysis simultaneously models  the RV and FWHM of the CCFs, to constrain the component of the RV signal that is caused by stellar activity. We consider a Gaussian process (GP) with shared hyperparameters between the RVs and FWHM, with the exception of the amplitudes. In the RVs, the model further assumes two Keplerian signals. Following \citet{faria_candidate_2022}, we subdivide the ESPRESSO data set into three subsets, ESP18, ESP19, and ESP21, and we consider possible RV offsets between these subsets. All priors are the same as those considered in \citet{faria_candidate_2022}.

The two signals with periods of 11.19 and 5.12 days are recovered with very high significance in the telluric-corrected RVs. When compared to the DRS RVs, the constraints on the orbital parameters of Proxima b are similar, but the semi-amplitude and eccentricity of Proxima d are found to be smaller, although within 1 sigma (Table\,\ref{table:paramProxima}). This decrease goes in the same direction as the result obtained with the template-matching RVs analyzed by \cite{faria_candidate_2022}, which uses the real RV content of each line but masks the telluric lines. We do not perform a direct comparison between the telluric-corrected CCF RVs and the template-matching RVs as the gain in photon noise and increase in parameters constraints arise from two different aspects\footnote{We also note that the template-matching analysis of \cite{faria_candidate_2022} provides the most constrained parameter values for the Proxima system.}. \\

\begin{figure}[h]
   \centering
   \includegraphics[width=\hsize]{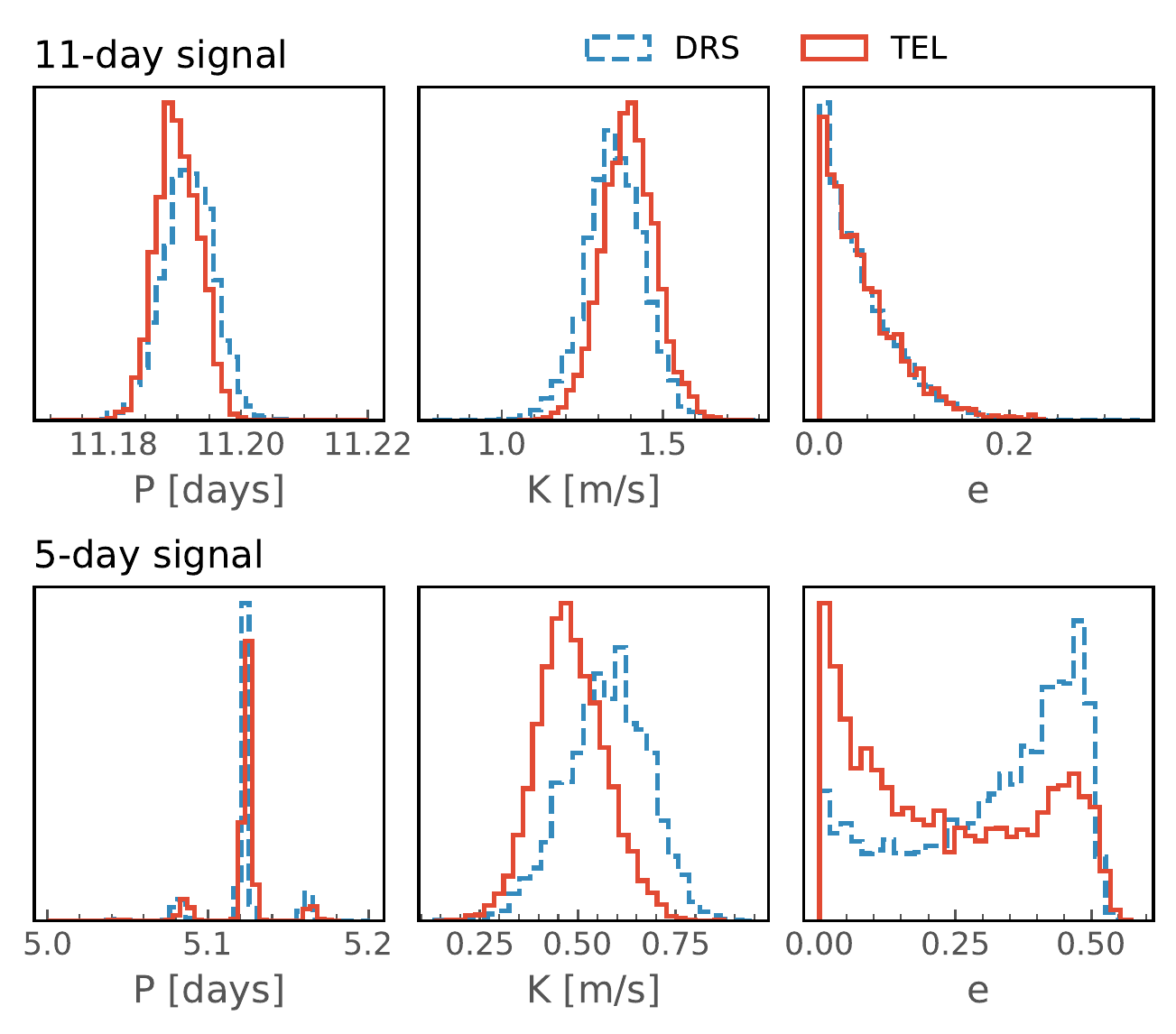}
      \caption{Posterior distributions for the orbital period, semi-amplitude, and eccentricity of the 11-day (top) and 5-day (bottom) signals. The dashed blue histograms correspond to the analysis of the DRS RVs, while the red histograms correspond to the analysis of the telluric-corrected RVs (TEL).}
         \label{FigVibStab}
   \end{figure}

\begin{table}
   \caption{Parameters of Proxima d from the analysis of DRS RVs \citep{faria_candidate_2022} and telluric-corrected RVs (TEL).}
   \label{table:paramProxima}
   \centering                          
   \begin{tabular}{c c c}
   \hline
       & DRS & TEL \\
       \hline
   Period [days] & 5.1214$^{+0.0015}_{-0.0019}$ & 5.1244$^{+0.0030}_{-0.0022}$\\
   Semi-amplitude [$\mathrm{m \cdot s^{-1}}$] & 0.65$^{+0.10}_{-0.10}$ &0.48$^{+0.10}_{-0.08}$ \\
   eccentricity  &0.37$^{+0.13}_{-0.23}$&0.20$^{+0.24}_{-0.16}$ \\
   M$_p$sin$i$ [M$_{\oplus}$] & 0.40$^{+0.06}_{-0.06}$&0.30$^{+0.05}_{-0.05}$  \\
   \hline
   \end{tabular}
   \end{table}
   
Considering the maximum-likelihood solution, the rms of the RV residuals reaches 34\,$\mathrm{cm \cdot s^{-1}}$ in the telluric-corrected RVs, compared to about 40\,$\mathrm{cm \cdot s^{-1}}$ in the DRS RVs. This improvement is also visible in each individual subset of the ESPRESSO data (Table \ref{table:rmsProxima}).

\begin{table}
   \caption{Residual rms for the full data set and individual subsets from the analysis of DRS RVs and telluric-corrected RVs (TEL).}
   \label{table:rmsProxima}
   \centering                          
   \begin{tabular}{c c c c c c}
   \hline
       & full data set & ESP18 & ESP19 & ESP21 \\
   \hline
   DRS & 40.4 & 38.9 & 47.7 & 38.7 & $\mathrm{cm \cdot s^{-1}}$ \\
   TEL &        34.8 & 34.3 & 45.9 & 30.1 & $\mathrm{cm \cdot s^{-1}}$ \\
   \hline                                   
   \end{tabular}
   \end{table}

Another comparison, which tries to include the full information in the posterior distributions, uses the so-called information, a measure of how much the prior is compressed to reach the posterior. Since the priors are the same in the analysis of the telluric-corrected RVs and DRS RVs, a larger compression means that the data provide more constraints on the model parameters. We obtain a value for the information of I=61 nats\footnote{One \emph{nat} is the information content of an event that has a probability equal to $e^{-1}$.} for the telluric-corrected RVs and I=59 nats for the DRS RVs. Although the values are close, they again reflect the improvement coming from using the telluric-corrected RVs.


\section{Application to the O$_2$ molecular bands}\label{Sec_O2}
The optical part of the telluric spectrum, in addition to the water bands discussed above, is dominated by four oxygen bands: $\delta$ ($\sim$579\,nm), $\gamma$ ($\sim$629\,nm), B ($\sim$688\,nm) and A ($\sim$765\,nm). From Fig.\,\ref{ESPRESSO_tell}, one can see that the B- and A-bands are dominated by saturated lines. There is no gain in correcting saturated or nearly saturated lines, which have an absorption depth above 90\%. Therefore, we consider continuing to mask these two bands for RV purposes but applying our telluric model to the $\gamma$-band, where the gain in spectral coverage, and thus RV content, is nonzero.\\

The physical parameters for $^{16}$O$_2$ are from HITRAN \citep{rothman_hitran2012_2013,gordon_hitran2016_2017}. We found that a Lorentzian profile was not sufficient to accurately model the oxygen lines, whereas a Voigt profile with a free thermal broadening for the line core was more efficient. As opposed to water vapor, oxygen is present higher up in the Earth's atmosphere, where the pressure broadening becomes negligible. We therefore added the temperature to the integrated oxygen column density and pressure, as a free parameter in the minimization fit. Equation\,\ref{T_nu_const} can thus be modified as follows:

\begin{eqnarray}
T(\nu) = e^{- IOC_{LOS} \cdot n_{O_2} \cdot \sum \sigma(\nu)  }
\label{T_nu_const_O2},
\end{eqnarray}  
where $n_{O_{2}} = 5.65 \cdot 10^{18}$ $\mathrm{molecules \cdot cm^{-3}}$ and $IOC$ is the integrated oxygen column density at the zenith, and is about 7\,km. On the contrary to the $IWV$, the $IOC_{LOS}$ varies mainly as a function of the airmass. This correlation explains the strong dependence of the RVs that are measured on O$_2$ lines on the airmass measured with HARPS \citep{figueira_evaluating_2010}.\\

\begin{figure}[h]
\resizebox{\hsize}{!}{\includegraphics{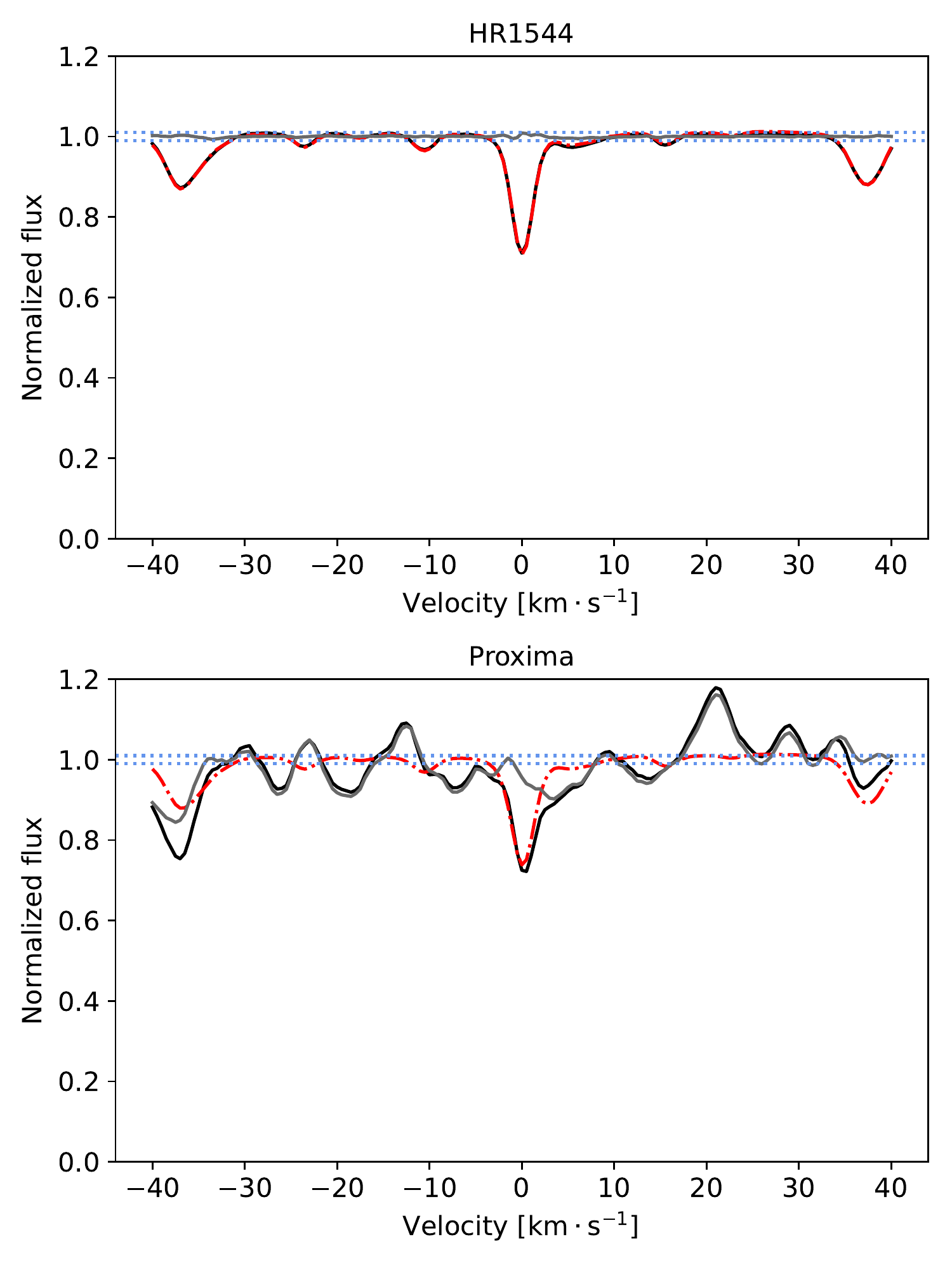}}
\caption{Master observed telluric CCF (in black), master modeled telluric CCF (in red, dashed) and master observed telluric-corrected CCF (in gray) for HR\,1544 (\textit{top}) and Proxima (\textit{bottom}). The CCFs are performed over the subset of selected O$_2$ lines in the $\gamma$-band. The two horizontal blue dotted lines encompass the 2\% residual P2V.}
\label{ccf_O2}
\end{figure}

Figure\,\ref{ccf_O2} shows the master observed telluric CCF, the master modeled telluric CCF, and the master observed telluric-corrected CCF, all obtained with the subset of O$_2$ lines in the $\gamma$-band for HR\,1544 and Proxima. In both cases, telluric residuals are below the 2\% P2V amplitude or within the continuum dispersion as in the case of water lines. This proves that the telluric model described here can be applied to different molecules.

\section{Conclusion}\label{Sec_concl}
The telluric correction presented here is based on a simple line-by-line telluric model, which requires the physical parameters of the lines (e.g., from the HITRAN database \citep{rothman_hitran2012_2013}), the sky conditions, and a detailed map of the instrument's spectral resolution as a function of the position on the detector. Then, the model is adjusted to the observations through a Levenberg-Marquardt minimization by building an average line profile over a subset of selected lines with strong absorptions. \\
\\
While this model can be applied to any molecules, our focus in this paper is the study of H$_2$O lines in the spectral range of ESPRESSO (3800-7800\,\AA). The water vapor lines are best represented by Lorentzian profiles, which are mainly dominated by the integrated water vapor that controls the line strength, but also by the pressure that controls the line width. We demonstrated that the telluric correction works for all stellar spectral types with residuals below a 2\,\% P2V amplitude. Similar performances have been obtained with other methods \citep[e.g.,][]{artigau_telluric-line_2014,smette_molecfit_2015,ulmer-moll_telluric_2019, leet_toward_2019,bedell_wobble_2019,cretignier_yarara_2021}, but our model does not require a telluric database, can be directly applied to a single observation, and is simple enough to be implemented in the DRS of state-of-the-art spectrographs to deliver high-fidelity spectra and RVs.\\
Telluric lines with absorption depths greater than 70-80\% exhibit residual structures with P-Cygni-like profiles. Several reasons can explain these structures, such as a lack of sophistication in our one-layer model (even if more complex models show similar results \citep{smette_molecfit_2015,allart_wasp-127b_2020}), inexact parameters in the HITRAN database, the presence of additional phenomena that deformed the line shape (e.g. winds), or systematic errors due to the noncommutative arithmetic between convolution and division. We note that Earth scientists use more complex line shapes to model the telluric lines but the level of complexity and number of parameters to take into account are incompatible with the simplicity that we aimed for here. \\
\\
The main goal of this telluric correction is to provide more precise RVs, unbiased by micro-telluric lines \citep{cunha_impact_2013}. The example of Tau Ceti showcased that micro-telluric lines can introduce systematics up to 58\,cm$\cdot$s$^{-1}$ P2V with an rms of 10\,cm$\cdot$s$^{-1}$. It is therefore critical to properly correct them when searching for Earth-mass planets. We also compared the RVs produced by the standard masks of the ESPRESSO DRS with different custom masks, testing the impact of the inclusion of a larger spectral region. To stick to the original idea of the ESPRESSO DRS, we kept these masks independent of the BERV and systemic velocities of the stars. Our best custom mask, \texttt{Orders77\_78}, while excluding orders 77 and 78 (where a couple of telluric water lines still have strong residuals), enables a gain of $\sim$500\,\AA\ (only $\sim$14\% of the spectral range is then excluded, compared to $\sim$27\% with the DRS mask) without introducing systematics. In the case of Proxima, this additional wavelength coverage leads to a gain of 25\,\% in photon noise or a factor of 1.78 in exposure time. This can lead to the detection of fainter RV signals but also a more efficient use of telescope time.\\
\\
We applied the same analysis as \cite{faria_candidate_2022} to the telluric-corrected CCF RVs of Proxima and we confirmed the presence of the planet Proxima b and planet candidate Proxima d. While the parameters of Proxima b are similar in both analyses, the semi-amplitude and eccentricity of Proxima d are smaller and their distributions go in the direction of the template matching analysis. 
Although the improvements brought by those two techniques converge to similar results for the parameters of planet candidate d, we want to note that those two methods do optimize completely different things. On one side, template matching uses the real RV content of each line and the most RV information possible, as in M dwarfs there is RV information everywhere, even in the pseudo-continuum. On the other side, the method proposed in this paper corrects for telluric absorption, and therefore a gain in RV information that leads to a better estimation of the planetary signal. We also want to note that improvements in the definition of the weights of each stellar line for the CCF technique are underway \citep{bourrier_rossiter-mclaughlin_2021}. \\
\\
We extended the application of our correction to the O$_2$ $\gamma$-band to display that our model can correct the telluric lines of molecules other than H$_2$O. In the case of the $\gamma$-band, the telluric residuals are within the requirements to perform extremely precise RV measurements, even if the spectral range freed remains modest ($\sim$80\,\AA).\\
\\
Adapting the model to other instruments, such as HARPS and HARPS-N instruments, will lead to a substantial increase in RV content. A gain in photon noise of 10\,\% could be reached for M5-type stars, and a $\sim$60\,cm$\cdot$s$^{-1}$ P2V micro-telluric amplitude could be removed. Furthermore, near-infrared high-resolution spectrographs, such as NIRPS, are the most important spectographs for applying a well-functioning telluric correction. At these wavelengths, H$_2$O is not the only main absorber, and CO$_2$, O$_2$, and CH$_4$ absorb on large spectral ranges, and with deep lines. It is, therefore, critical to optimize the telluric correction to facilitate the maximum extraction of RV information.\\
\\
The impact of the telluric correction for all the stars observed by the ESPRESSO consortium still needs to be investigated in detail but we expect that it will lead to more precise and less biased discoveries. The impact on other science cases, such as stellar activity indicators, atmospheric characterization, or the measurement of the fine structure constant, still has to be studied. Removing micro-telluric lines is potentially very important for measurements of the fine structure constant with ESPRESSO: micro-telluric lines which fall on top of the relevant transitions in the quasar spectrum could spoil the measurement if not removed or recognized. Therefore, we can be optimistic regarding the benefits of applying our correction methods to different science cases. \\


\begin{acknowledgements}
We thank the referee for their useful comments that improve the overall paper quality. R.A. thanks Etienne Artigau for useful discussions on the behavior of the telluric lines, and the comparison with his telluric correction for the SPIRou instrument. R.A. thanks Alain Smette and Martin Turbet for useful discussions on the telluric line behavior and line shape. R.A. is a Trottier Postdoctoral Fellow and acknowledges support from the Trottier Family Foundation. This work was supported in part through a grant from FRQNT. This work has been carried out within the framework of the National Centre of Competence in Research PlanetS supported by the Swiss National Science Foundation. The authors acknowledge the financial support of the SNSF. This project has received funding from the European Research Council (ERC) under the European Union’s Horizon 2020 research and innovation program (grant agreement SCORE No 851555). A.M.S acknowledges support from the Fundação para a Ciência e a Tecnologia (FCT) through the Fellowship 2020.05387.BD. and POCH/FSE (EC). This  work  was supported  by  FCT  - Fundação para   a   Ciência   e   a   Tecnologia   through national   funds   and   by   FEDER through  COMPETE2020  - Programa  Operacional  Competitividade  e  Inter-nacionalização by  these  grants: UID/FIS/04434/2019;   UIDB/04434/2020; UIDP/04434/2020;    PTDC/FIS-AST/32113/2017   \& POCI-01-0145-FEDER-032113; PTDC/FIS-AST/28953/2017    \& POCI-01-0145-FEDER-028953; PTDC/FIS-AST/28987/2017   \& POCI-01-0145-FEDER-028987. J.I.G.H., R.R., and A.S.M. acknowledge financial support from the Spanish Ministry of Science and Innovation (MICIN) project PID2020-117493GB-I00. J.I.G.H. and R.R. also acknowledge financial support from the Government of the Canary Islands project ProID2020010129. A.S.M. acknowledges financial support from the Spanish Ministry of Science and Innovation (MICINN) under 2018 Juan de la Cierva program IJC2018-035229-I. N.J.N. acknowledges support from the following projects: UIDB/04434/2020 \& UIDP/04434/2020, CERN/FIS-PAR/0037/2019, PTDC/FIS-OUT/29048/2017,  COMPETE2020: POCI-01-0145-FEDER-028987 \& FCT: PTDC/FIS-AST/28987/2017, PTDC/FIS-AST/0054/2022. D.M. is also supported by the INFN PD51 INDARK grant.

\end{acknowledgements}
\bibliographystyle{aa}
\bibliography{bib}

\begin{thebibliography}{42}
\expandafter\ifx\csname natexlab\endcsname\relax\def\natexlab#1{#1}\fi

\bibitem[{Allart {et~al.}(2017)Allart, Lovis, Pino, Wyttenbach, Ehrenreich, \&
  Pepe}]{allart_search_2017}
Allart, R., Lovis, C., Pino, L., {et~al.} 2017, Astronomy and Astrophysics,
  606, A144

\bibitem[{Allart {et~al.}(2020)Allart, Pino, Lovis, Sousa, Casasayas-Barris,
  Zapatero~Osorio, Cretignier, Palle, Pepe, Cristiani, Rebolo, Santos, Borsa,
  Bourrier, Demangeon, Ehrenreich, Lavie, Lendl, Lillo-Box, Micela, Oshagh,
  Sozzetti, Tabernero, Adibekyan, Allende~Prieto, Alibert, Amate, Benz, Bouchy,
  Cabral, Dekker, D'Odorico, Di~Marcantonio, Dumusque, Figueira, Genova~Santos,
  Gonz{\'a}lez~Hern{\'a}ndez, Lo~Curto, Manescau, Martins, M{\'e}gevand,
  Mehner, Molaro, Nunes, Poretti, Riva, Su{\'a}rez~Mascare{\~n}o, Udry, \&
  Zerbi}]{allart_wasp-127b_2020}
Allart, R., Pino, L., Lovis, C., {et~al.} 2020, Astronomy and Astrophysics,
  644, A155

\bibitem[{Anglada-Escud{\'e} {et~al.}(2016)Anglada-Escud{\'e}, Amado, Barnes,
  Berdi{\~n}as, Butler, Coleman, de~La~Cueva, Dreizler, Endl, Giesers, Jeffers,
  Jenkins, Jones, Kiraga, K{\"u}rster, L{\'o}pez-Gonz{\'a}lez, Marvin, Morales,
  Morin, Nelson, Ortiz, Ofir, Paardekooper, Reiners, Rodr{\'i}guez,
  Rodr{\'i}guez-L{\'o}pez, Sarmiento, Strachan, Tsapras, Tuomi, \&
  Zechmeister}]{anglada-escude_terrestrial_2016}
Anglada-Escud{\'e}, G., Amado, P.~J., Barnes, J., {et~al.} 2016, Nature, 536,
  437, aDS Bibcode: 2016Natur.536..437A

\bibitem[{Anglada-Escud{\'e} \& Butler(2012)}]{anglada-escude_harps-terra_2012}
Anglada-Escud{\'e}, G. \& Butler, R.~P. 2012, The Astrophysical Journal
  Supplement Series, 200, 15, aDS Bibcode: 2012ApJS..200...15A

\bibitem[{Artigau {et~al.}(2014)Artigau, Astudillo-Defru, Delfosse, Bouchy,
  Bonfils, Lovis, Pepe, Moutou, Donati, Doyon, \&
  Malo}]{artigau_telluric-line_2014}
Artigau, {\'E}., Astudillo-Defru, N., Delfosse, X., {et~al.} 2014, 9149,
  914905, conference Name: Observatory Operations: Strategies, Processes, and
  Systems V Place: eprint: arXiv:1406.6927 ADS Bibcode: 2014SPIE.9149E..05A

\bibitem[{Artigau {et~al.}(2021)Artigau, H{\'e}brard, Cadieux, Vandal, Cook,
  Doyon, Gagn{\'e}, Moutou, Martioli, Frasca, Jahandar, Lafreni{\`e}re, Malo,
  Donati, Cort{\'e}s-Zuleta, Boisse, Delfosse, Carmona, Fouqu{\'e}, Morin,
  Rowe, Marino, Papini, Ciardi, Lund, Martins, Pelletier, Arnold, Bouchy,
  Forveille, Santos, Bonfils, Figueira, Fausnaugh, Ricker, Latham, Seager,
  Winn, Jenkins, Ting, Torres, \& Gomes~da Silva}]{artigau_toi-1278_2021}
Artigau, {\'E}., H{\'e}brard, G., Cadieux, C., {et~al.} 2021, The Astronomical
  Journal, 162, 144, aDS Bibcode: 2021AJ....162..144A

\bibitem[{Astudillo-Defru {et~al.}(2015)Astudillo-Defru, Bonfils, Delfosse,
  S{\'e}gransan, Forveille, Bouchy, Gillon, Lovis, Mayor, Neves, Pepe, Perrier,
  Queloz, Rojo, Santos, \& Udry}]{astudillo-defru_harps_2015}
Astudillo-Defru, N., Bonfils, X., Delfosse, X., {et~al.} 2015, Astronomy and
  Astrophysics, 575, A119

\bibitem[{Baker {et~al.}(2020)Baker, Blake, \& Reiners}]{baker_iag_2020}
Baker, A.~D., Blake, C.~H., \& Reiners, A. 2020, The Astrophysical Journal
  Supplement Series, 247, 24, aDS Bibcode: 2020ApJS..247...24B

\bibitem[{Baranne {et~al.}(1996)Baranne, Queloz, Mayor, Adrianzyk, Knispel,
  Kohler, Lacroix, Meunier, Rimbaud, \& Vin}]{baranne_elodie_1996}
Baranne, A., Queloz, D., Mayor, M., {et~al.} 1996, Astronomy and Astrophysics
  Supplement Series, 119, 373

\bibitem[{Bedell {et~al.}(2019)Bedell, Hogg, Foreman-Mackey, Montet, \&
  Luger}]{bedell_wobble_2019}
Bedell, M., Hogg, D.~W., Foreman-Mackey, D., Montet, B.~T., \& Luger, R. 2019,
  The Astronomical Journal, 158, 164, aDS Bibcode: 2019AJ....158..164B

\bibitem[{Bertaux {et~al.}(2014)Bertaux, Lallement, Ferron, Boonne, \&
  Bodichon}]{bertaux_tapas_2014}
Bertaux, J.~L., Lallement, R., Ferron, S., Boonne, C., \& Bodichon, R. 2014,
  Astronomy \&amp; Astrophysics, Volume 564, id.A46,
  {\textless}NUMPAGES{\textgreater}12{\textless}/NUMPAGES{\textgreater} pp.,
  564, A46

\bibitem[{Bourrier {et~al.}(2021)Bourrier, Lovis, Cretignier, Allart, Dumusque,
  Delisle, Deline, Sousa, Adibekyan, Alibert, Barros, Borsa, Cristiani,
  Demangeon, Ehrenreich, Figueira, Gonz{\'a}lez~Hern{\'a}ndez, Lendl,
  Lillo-Box, Lo~Curto, Di~Marcantonio, Martins, M{\'e}gevand, Mehner, Micela,
  Molaro, Oshagh, Palle, Pepe, Poretti, Rebolo, Santos, Scandariato, Seidel,
  Sozzetti, Su{\'a}rez~Mascare{\~n}o, \&
  Zapatero~Osorio}]{bourrier_rossiter-mclaughlin_2021}
Bourrier, V., Lovis, C., Cretignier, M., {et~al.} 2021, Astronomy \&amp;
  Astrophysics, Volume 654, id.A152,
  {\textless}NUMPAGES{\textgreater}19{\textless}/NUMPAGES{\textgreater} pp.,
  654, A152

\bibitem[{Cretignier {et~al.}(2021)Cretignier, Dumusque, Hara, \&
  Pepe}]{cretignier_yarara_2021}
Cretignier, M., Dumusque, X., Hara, N.~C., \& Pepe, F. 2021, Astronomy and
  Astrophysics, 653, A43

\bibitem[{Cunha {et~al.}(2013)Cunha, Figueira, Santos, Lovis, \&
  Bou{\'e}}]{cunha_impact_2013}
Cunha, D., Figueira, P., Santos, N.~C., Lovis, C., \& Bou{\'e}, G. 2013,
  Astronomy and Astrophysics, 550, A75

\bibitem[{Damasso {et~al.}(2020)Damasso, Del~Sordo, Anglada-Escud{\'e},
  Giacobbe, Sozzetti, Morbidelli, Pojmanski, Barbato, Butler, Jones, Hambsch,
  Jenkins, L{\'o}pez-Gonz{\'a}lez, Morales, Pe{\~n}a~Rojas,
  Rodr{\'i}guez-L{\'o}pez, Rodr{\'i}guez, Amado, Anglada, Feng, \&
  G{\'o}mez}]{damasso_low-mass_2020}
Damasso, M., Del~Sordo, F., Anglada-Escud{\'e}, G., {et~al.} 2020, Science
  Advances, 6, eaax7467

\bibitem[{Donati {et~al.}(2020)Donati, Kouach, Moutou, Doyon, Delfosse,
  Artigau, Baratchart, Lacombe, Barrick, H{\'e}brard, Bouchy, Saddlemyer,
  Par{\`e}s, Rabou, Micheau, Dolon, Reshetov, Challita, Carmona, Striebig,
  Thibault, Martioli, Cook, Fouqu{\'e}, Vermeulen, Wang, Arnold, Pepe, Boisse,
  Figueira, Bouvier, Ray, Feugeade, Morin, Alencar, Hobson, Castilho, Udry,
  Santos, Hernandez, Benedict, Vall{\'e}e, Gallou, Dupieux, Larrieu, Perruchot,
  Sottile, Moreau, Usher, Baril, Wildi, Chazelas, Malo, Bonfils, Loop, Kerley,
  Wevers, Dunn, Pazder, Macdonald, Dubois, Carri{\'e}, Valentin, Henault, Yan,
  \& Steinmetz}]{donati_spirou_2020}
Donati, J.~F., Kouach, D., Moutou, C., {et~al.} 2020, Monthly Notices of the
  Royal Astronomical Society, 498, 5684, aDS Bibcode: 2020MNRAS.498.5684D

\bibitem[{Dumusque {et~al.}(2011)Dumusque, Udry, Lovis, Santos, \&
  Monteiro}]{dumusque_planetary_2011}
Dumusque, X., Udry, S., Lovis, C., Santos, N.~C., \& Monteiro, M. J. P. F.~G.
  2011, Astronomy and Astrophysics, 525, A140

\bibitem[{Faria {et~al.}(2022)Faria, Su{\'a}rez~Mascare{\~n}o, Figueira, Silva,
  Damasso, Demangeon, Pepe, Santos, Rebolo, Cristiani, Adibekyan, Alibert,
  Allart, Barros, Cabral, D'Odorico, Di~Marcantonio, Dumusque, Ehrenreich,
  Gonz{\'a}lez~Hern{\'a}ndez, Hara, Lillo-Box, Lo~Curto, Lovis, Martins,
  M{\'e}gevand, Mehner, Micela, Molaro, Nunes, Pall{\'e}, Poretti, Sousa,
  Sozzetti, Tabernero, Udry, \& Zapatero~Osorio}]{faria_candidate_2022}
Faria, J.~P., Su{\'a}rez~Mascare{\~n}o, A., Figueira, P., {et~al.} 2022,
  Astronomy \&amp; Astrophysics, Volume 658, id.A115,
  {\textless}NUMPAGES{\textgreater}16{\textless}/NUMPAGES{\textgreater} pp.,
  658, A115

\bibitem[{Feng {et~al.}(2017)Feng, Tuomi, Jones, Barnes, Anglada-Escud{\'e},
  Vogt, \& Butler}]{feng_color_2017}
Feng, F., Tuomi, M., Jones, H. R.~A., {et~al.} 2017, The Astronomical Journal,
  154, 135

\bibitem[{Figueira {et~al.}(2010)Figueira, Pepe, Lovis, \&
  Mayor}]{figueira_evaluating_2010}
Figueira, P., Pepe, F., Lovis, C., \& Mayor, M. 2010, Astronomy and
  Astrophysics, 515, A106

\bibitem[{Gordon {et~al.}(2017)Gordon, Rothman, Hill, Kochanov, Tan, Bernath,
  Birk, Boudon, Campargue, Chance, Drouin, Flaud, Gamache, Hodges, Jacquemart,
  Perevalov, Perrin, Shine, Smith, Tennyson, Toon, Tran, Tyuterev, Barbe,
  Cs{\'a}sz{\'a}r, Devi, Furtenbacher, Harrison, Hartmann, Jolly, Johnson,
  Karman, Kleiner, Kyuberis, Loos, Lyulin, Massie, Mikhailenko, Moazzen-Ahmadi,
  M{\"u}ller, Naumenko, Nikitin, Polyansky, Rey, Rotger, Sharpe, Sung,
  Starikova, Tashkun, Auwera, Wagner, Wilzewski, Wcis{\l }o, Yu, \&
  Zak}]{gordon_hitran2016_2017}
Gordon, I.~E., Rothman, L.~S., Hill, C., {et~al.} 2017, Journal of Quantitative
  Spectroscopy and Radiative Transfer, 203, 3, aDS Bibcode: 2017JQSRT.203....3G

\bibitem[{Gullikson {et~al.}(2014)Gullikson, Dodson-Robinson, \&
  Kraus}]{gullikson_correcting_2014}
Gullikson, K., Dodson-Robinson, S., \& Kraus, A. 2014, The Astronomical
  Journal, 148, 53, aDS Bibcode: 2014AJ....148...53G

\bibitem[{Jurgenson {et~al.}(2016)Jurgenson, Fischer, McCracken, Sawyer,
  Szymkowiak, Davis, Muller, \& Santoro}]{jurgenson_expres_2016}
Jurgenson, C., Fischer, D., McCracken, T., {et~al.} 2016, 9908, 99086T,
  conference Name: Ground-based and Airborne Instrumentation for Astronomy VI
  ISBN: 9781510601956 Place: eprint: arXiv:1606.04413

\bibitem[{Kausch {et~al.}(2015)Kausch, Noll, Smette, Kimeswenger, Barden,
  Szyszka, Jones, Sana, Horst, \& Kerber}]{kausch_molecfit_2015}
Kausch, W., Noll, S., Smette, A., {et~al.} 2015, Astronomy and Astrophysics,
  576, A78

\bibitem[{Kerber {et~al.}(2012)Kerber, Rose, van~den Ancker, \&
  Querel}]{kerber_monitoring_2012}
Kerber, F., Rose, T., van~den Ancker, M., \& Querel, R.~R. 2012, The Messenger,
  148, 9

\bibitem[{Leet {et~al.}(2019)Leet, Fischer, \& Valenti}]{leet_toward_2019}
Leet, C., Fischer, D.~A., \& Valenti, J.~A. 2019, The Astronomical Journal,
  157, 187, aDS Bibcode: 2019AJ....157..187L

\bibitem[{Li {et~al.}(2018)Li, Blake, Nidever, \& Halverson}]{li_temporal_2018}
Li, D., Blake, C.~H., Nidever, D., \& Halverson, S.~P. 2018, Publications of
  the Astronomical Society of the Pacific, 130, 014501, aDS Bibcode:
  2018PASP..130a4501L

\bibitem[{Lisogorskyi {et~al.}(2019)Lisogorskyi, Jones, \&
  Feng}]{lisogorskyi_activity_2019}
Lisogorskyi, M., Jones, H. R.~A., \& Feng, F. 2019, Monthly Notices of the
  Royal Astronomical Society, 485, 4804, aDS Bibcode: 2019MNRAS.485.4804L

\bibitem[{Mazeh {et~al.}(2007)Mazeh, Tamuz, \& Zucker}]{mazeh_sys-rem_2007}
Mazeh, T., Tamuz, O., \& Zucker, S. 2007, 366, 119, conference Name: Transiting
  Extrapolar Planets Workshop Place: eprint: arXiv:astro-ph/0612418 ADS
  Bibcode: 2007ASPC..366..119M

\bibitem[{Pepe {et~al.}(2021)Pepe, Cristiani, Rebolo, Santos, Dekker, Cabral,
  Di~Marcantonio, Figueira, Lo~Curto, Lovis, Mayor, M{\'e}gevand, Molaro, Riva,
  Zapatero~Osorio, Amate, Manescau, Pasquini, Zerbi, Adibekyan, Abreu,
  Affolter, Alibert, Aliverti, Allart, Allende~Prieto, {\'A}lvarez, Alves,
  Avila, Baldini, Bandy, Barros, Benz, Bianco, Borsa, Bourrier, Bouchy, Broeg,
  Calderone, Cirami, Coelho, Conconi, Coretti, Cumani, Cupani, D'Odorico,
  Damasso, Deiries, Delabre, Demangeon, Dumusque, Ehrenreich, Faria, Fragoso,
  Genolet, Genoni, G{\'e}nova~Santos, Gonz{\'a}lez~Hern{\'a}ndez, Hughes,
  Iwert, Kerber, Knudstrup, Landoni, Lavie, Lillo-Box, Lizon, Maire, Martins,
  Mehner, Micela, Modigliani, Monteiro, Monteiro, Moschetti, Murphy, Nunes,
  Oggioni, Oliveira, Oshagh, Pall{\'e}, Pariani, Poretti, Rasilla,
  Rebord{\~a}o, Redaelli, Santana~Tschudi, Santin, Santos, S{\'e}gransan,
  Schmidt, Segovia, Sosnowska, Sozzetti, Sousa, Span{\`o},
  Su{\'a}rez~Mascare{\~n}o, Tabernero, Tenegi, Udry, \&
  Zanutta}]{pepe_espresso_2021}
Pepe, F., Cristiani, S., Rebolo, R., {et~al.} 2021, Astronomy \&amp;
  Astrophysics, Volume 645, id.A96,
  {\textless}NUMPAGES{\textgreater}26{\textless}/NUMPAGES{\textgreater} pp.,
  645, A96

\bibitem[{Pepe {et~al.}(2002)Pepe, Mayor, Galland, Naef, Queloz, Santos, Udry,
  \& Burnet}]{pepe_coralie_2002}
Pepe, F., Mayor, M., Galland, F., {et~al.} 2002, Astronomy and Astrophysics,
  388, 632

\bibitem[{Quirrenbach {et~al.}(2014)Quirrenbach, Amado, Caballero, Mundt,
  Reiners, Ribas, Seifert, Abril, Aceituno, Alonso-Floriano, Ammler-von Eiff,
  Antona~Jim{\'e}nez, Anwand-Heerwart, Azzaro, Bauer, Barrado, Becerril,
  B{\'e}jar, Ben{\'i}tez, Berdi{\~n}as, C{\'a}rdenas, Casal, Claret,
  Colom{\'e}, Cort{\'e}s-Contreras, Czesla, Doellinger, Dreizler, Feiz,
  Fern{\'a}ndez, Galad{\'i}, G{\'a}lvez-Ortiz, Garc{\'i}a-Piquer,
  Garc{\'i}a-Vargas, Garrido, Gesa, G{\'o}mez~Galera, Gonz{\'a}lez~{\'A}lvarez,
  Gonz{\'a}lez~Hern{\'a}ndez, Gr{\"o}zinger, Gu{\`a}rdia, Guenther, de~Guindos,
  Guti{\'e}rrez-Soto, Hagen, Hatzes, Hauschildt, Helmling, Henning, Hermann,
  Hern{\'a}ndez~Casta{\~n}o, Herrero, Hidalgo, Holgado, Huber, Huber, Jeffers,
  Joergens, de~Juan, Kehr, Klein, K{\"u}rster, Lamert, Lalitha, Laun, Lemke,
  Lenzen, L{\'o}pez~del Fresno, L{\'o}pez~Mart{\'i}, L{\'o}pez-Santiago, Mall,
  Mandel, Mart{\'i}n, Mart{\'i}n-Ruiz, Mart{\'i}nez-Rodr{\'i}guez, Marvin,
  Mathar, Mirabet, Montes, Morales~Mu{\~n}oz, Moya, Naranjo, Ofir, Oreiro,
  Pall{\'e}, Panduro, Passegger, P{\'e}rez-Calpena, P{\'e}rez~Medialdea,
  Perger, Pluto, Ram{\'o}n, Rebolo, Redondo, Reffert, Reinhardt, Rhode, Rix,
  Rodler, Rodr{\'i}guez, Rodr{\'i}guez-L{\'o}pez, Rodr{\'i}guez-P{\'e}rez,
  Rohloff, Rosich, S{\'a}nchez-Blanco, S{\'a}nchez~Carrasco, Sanz-Forcada,
  Sarmiento, Sch{\"a}fer, Schiller, Schmidt, Schmitt, Solano, Stahl, Storz,
  St{\"u}rmer, Su{\'a}rez, Ulbrich, Veredas, Wagner, Winkler, Zapatero~Osorio,
  Zechmeister, Abell{\'a}n~de Paco, Anglada-Escud{\'e}, del Burgo, Klutsch,
  Lizon, L{\'o}pez-Morales, Morales, Perryman, Tulloch, \&
  Xu}]{quirrenbach_carmenes_2014}
Quirrenbach, A., Amado, P.~J., Caballero, J.~A., {et~al.} 2014, in , 91471F

\bibitem[{Rothman {et~al.}(2013)Rothman, Gordon, Babikov, Barbe, Chris~Benner,
  Bernath, Birk, Bizzocchi, Boudon, Brown, Campargue, Chance, Cohen, Coudert,
  Devi, Drouin, Fayt, Flaud, Gamache, Harrison, Hartmann, Hill, Hodges,
  Jacquemart, Jolly, Lamouroux, Le~Roy, Li, Long, Lyulin, Mackie, Massie,
  Mikhailenko, M{\"u}ller, Naumenko, Nikitin, Orphal, Perevalov, Perrin,
  Polovtseva, Richard, Smith, Starikova, Sung, Tashkun, Tennyson, Toon,
  Tyuterev, \& Wagner}]{rothman_hitran2012_2013}
Rothman, L.~S., Gordon, I.~E., Babikov, Y., {et~al.} 2013, Journal of
  Quantitative Spectroscopy and Radiative Transfer, 130, 4, aDS Bibcode:
  2013JQSRT.130....4R

\bibitem[{Sameshima {et~al.}(2018)Sameshima, Matsunaga, Kobayashi, Kawakita,
  Hamano, Ikeda, Kondo, Fukue, Taniguchi, Mizumoto, Arai, Otsubo, Takenaka,
  Watase, Asano, Yasui, Izumi, \& Yoshikawa}]{sameshima_correction_2018}
Sameshima, H., Matsunaga, N., Kobayashi, N., {et~al.} 2018, Publications of the
  Astronomical Society of the Pacific, 130, 074502, aDS Bibcode:
  2018PASP..130g4502S

\bibitem[{Smette {et~al.}(2015)Smette, Sana, Noll, Horst, Kausch, Kimeswenger,
  Barden, Szyszka, Jones, Gallenne, Vinther, Ballester, \&
  Taylor}]{smette_molecfit_2015}
Smette, A., Sana, H., Noll, S., {et~al.} 2015, Astronomy and Astrophysics, 576,
  A77

\bibitem[{Snellen {et~al.}(2010)Snellen, de~Kok, de~Mooij, \&
  Albrecht}]{snellen_orbital_2010}
Snellen, I. A.~G., de~Kok, R.~J., de~Mooij, E. J.~W., \& Albrecht, S. 2010,
  Nature, 465, 1049, aDS Bibcode: 2010Natur.465.1049S

\bibitem[{Su{\'a}rez~Mascare{\~n}o {et~al.}(2020)Su{\'a}rez~Mascare{\~n}o,
  Faria, Figueira, Lovis, Damasso, Gonz{\'a}lez~Hern{\'a}ndez, Rebolo,
  Cristiani, Pepe, Santos, Zapatero~Osorio, Adibekyan, Hojjatpanah, Sozzetti,
  Murgas, Abreu, Affolter, Alibert, Aliverti, Allart, Allende~Prieto, Alves,
  Amate, Avila, Baldini, Bandi, Barros, Bianco, Benz, Bouchy, Broeng, Cabral,
  Calderone, Cirami, Coelho, Conconi, Coretti, Cumani, Cupani, D'Odorico,
  Deiries, Delabre, Di~Marcantonio, Dumusque, Ehrenreich, Fragoso, Genolet,
  Genoni, G{\'e}nova~Santos, Hughes, Iwert, Kerber, Knusdstrup, Landoni, Lavie,
  Lillo-Box, Lizon, Lo~Curto, Maire, Manescau, Martins, M{\'e}gevand, Mehner,
  Micela, Modigliani, Molaro, Monteiro, Monteiro, Moschetti, Mueller, Nunes,
  Oggioni, Oliveira, Pall{\'e}, Pariani, Pasquini, Poretti, Rasilla, Redaelli,
  Riva, Santana~Tschudi, Santin, Santos, Segovia, Sosnowska, Sousa, Span{\`o},
  Tenegi, Udry, Zanutta, \& Zerbi}]{suarez_mascareno_revisiting_2020}
Su{\'a}rez~Mascare{\~n}o, A., Faria, J.~P., Figueira, P., {et~al.} 2020,
  Astronomy and Astrophysics, 639, A77

\bibitem[{Ulmer-Moll {et~al.}(2019)Ulmer-Moll, Figueira, Neal, Santos, \&
  Bonnefoy}]{ulmer-moll_telluric_2019}
Ulmer-Moll, S., Figueira, P., Neal, J.~J., Santos, N.~C., \& Bonnefoy, M. 2019,
  Astronomy and Astrophysics, 621, A79

\bibitem[{Vidal-Madjar {et~al.}(1986)Vidal-Madjar, Ferlet, Gry, \&
  Lallement}]{vidal-madjar_deuterium_1986}
Vidal-Madjar, A., Ferlet, R., Gry, C., \& Lallement, R. 1986, Astronomy and
  Astrophysics, Vol. 155, p. 407-412 (1986), 155, 407

\bibitem[{Wildi {et~al.}(2017)Wildi, Blind, Reshetov, Hernandez, Genolet,
  Conod, Sordet, Segovilla, Rasilla, Brousseau, Thibault, Delabre, Bandy,
  Sarajlic, Cabral, Bovay, Vall{\'e}e, Bouchy, Doyon, Artigau, Pepe, Hagelberg,
  Melo, Delfosse, Figueira, Santos, Gonz{\'a}lez~Hern{\'a}ndez, de~Medeiros,
  Rebolo, Broeg, Benz, Boisse, Malo, K{\"a}ufl, \&
  Saddlemyer}]{wildi_nirps_2017}
Wildi, F., Blind, N., Reshetov, V., {et~al.} 2017, 10400, 1040018, conference
  Name: Society of Photo-Optical Instrumentation Engineers (SPIE) Conference
  Series ADS Bibcode: 2017SPIE10400E..18W

\bibitem[{Xuesong~Wang {et~al.}(2019)Xuesong~Wang, Wright, Bender, Howard,
  Isaacson, Veyette, \& Muirhead}]{xuesong_wang_effects_2019}
Xuesong~Wang, S., Wright, J.~T., Bender, C., {et~al.} 2019, The Astronomical
  Journal, 158, 216, aDS Bibcode: 2019AJ....158..216X

\bibitem[{Zechmeister {et~al.}(2018)Zechmeister, Reiners, Amado, Azzaro, Bauer,
  B{\'e}jar, Caballero, Guenther, Hagen, Jeffers, Kaminski, K{\"u}rster,
  Launhardt, Montes, Morales, Quirrenbach, Reffert, Ribas, Seifert, Tal-Or, \&
  Wolthoff}]{zechmeister_spectrum_2018}
Zechmeister, M., Reiners, A., Amado, P.~J., {et~al.} 2018, Astronomy and
  Astrophysics, 609, A12

\end{thebibliography}

\begin{appendix}
\section{Coefficient of the resolution parametrization for the HR1x1 mode. }\label{AppendixA}

\begin{table}[h]
\centering
\caption{Parameters of Equation\,\ref{eq_resolution}}
\begin{tabular}{lcccc}
\hline
Coefficient      & Blue detector, slice 1 & Blue detector, slice 2 & Red detector, slice 1 & Red detector, slice 2 \\ 
\hline
\textit{a} &        6.71                               &      4.99                            & 21.27                  & 19.99 \\ 
\textit{b} &        160.43                           &      -98.59                        & 1212.60              & 866.67 \\ 
\textit{c} &        0.01                               &      0.05                            & -0.11                  &  -0.12 \\ 
\textit{d} &        -7$\cdot$10$^{-4}$      &      -9$\cdot$10$^{-4}$  & -0.001                 &  -8$\cdot$10$^{-4}$\\ 
\textit{e} &        -2.31                             &      -0.68                          & -3.2                     & -1.41 \\ 
\textit{f}  &        -2$\cdot$10$^{-8}$      &      -2$\cdot$10$^{-8}$  & -3$\cdot$10$^{-8}$ &  -3$\cdot$10$^{-8}$\\ 
\textit{g} &        119652                          &      127996                      & 27394                   &  41851\\ 
\hline
\end{tabular}
\label{Param_res}
\end{table}

\section{List of the 20 water telluric lines used in the subset and the telluric CCF computation }\label{AppendixB}

\begin{table}[h]
\centering
\caption{Wavelength position in vacuum and intensity of the 20 water lines used in the subset.}
\begin{tabular}{lcccc}
\hline
$\lambda$ [\AA]      & $S_{ij}$ [$\mathrm{cm^{-1} / (molecule \cdot cm^{-2})}$] \\ 
\hline
7179.34 & 1.65$\cdot$10$^{-23}$                               \\
7183.51 & 2.42$\cdot$10$^{-23}$ \\
7189.38 & 3.39$\cdot$10$^{-23}$                          \\ 
7193.47 & 3.84$\cdot$10$^{-23}$                             \\ 
7195.74 & 1.34$\cdot$10$^{-23}$          \\ 
7203.18 & 3.27$\cdot$10$^{-23}$                      \\ 
7206.28 & 2.62$\cdot$10$^{-23}$          \\ 
7225.61 & 1.22$\cdot$10$^{-23}$                      \\ 
7242.63 & 2.05$\cdot$10$^{-23}$ \\
7245.71 & 2.80$\cdot$10$^{-23}$ \\
7247.67 & 1.22$\cdot$10$^{-23}$ \\
7254.38 & 2.36$\cdot$10$^{-23}$ \\
7267.59 & 3.09$\cdot$10$^{-23}$ \\
7274.96 & 2.67$\cdot$10$^{-23}$ \\
7277.41 & 1.41$\cdot$10$^{-23}$ \\
7279.39 & 2.39$\cdot$10$^{-23}$ \\
7289.39 & 1.51$\cdot$10$^{-23}$ \\
7292.39 & 2.14$\cdot$10$^{-23}$ \\
7306.21 & 1.44$\cdot$10$^{-23}$ \\
7311.53 & 1.12$\cdot$10$^{-23}$ \\
\hline
\end{tabular}
\label{Param_res}
\end{table}

\section{Masks used}\label{AppendixC}
\begin{figure*}[h]

\includegraphics[width=\textwidth]{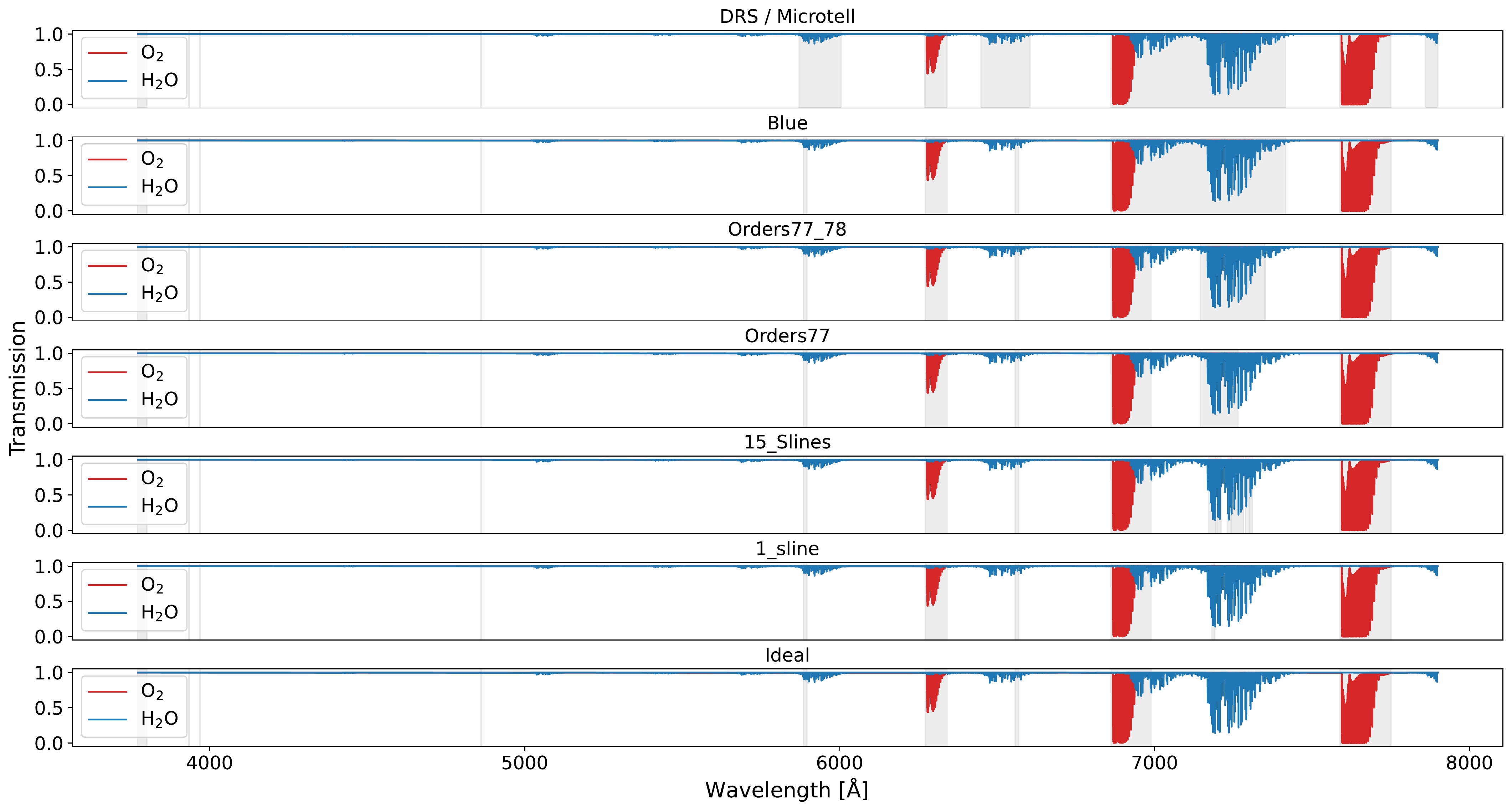}
\adjustbox{minipage=\textwidth,left}{\caption{Telluric spectrum in the range of ESPRESSO. In blue, the H$_2$O lines, and in red the O$_2$ lines. The different masks are shown in the panels and are ordered as a function of the spectral coverage included in the mask. In gray are the regions excluded from the mask.}\label{mask_applied}}
\end{figure*}

\begin{figure*}[h]
\includegraphics[width=\textwidth]{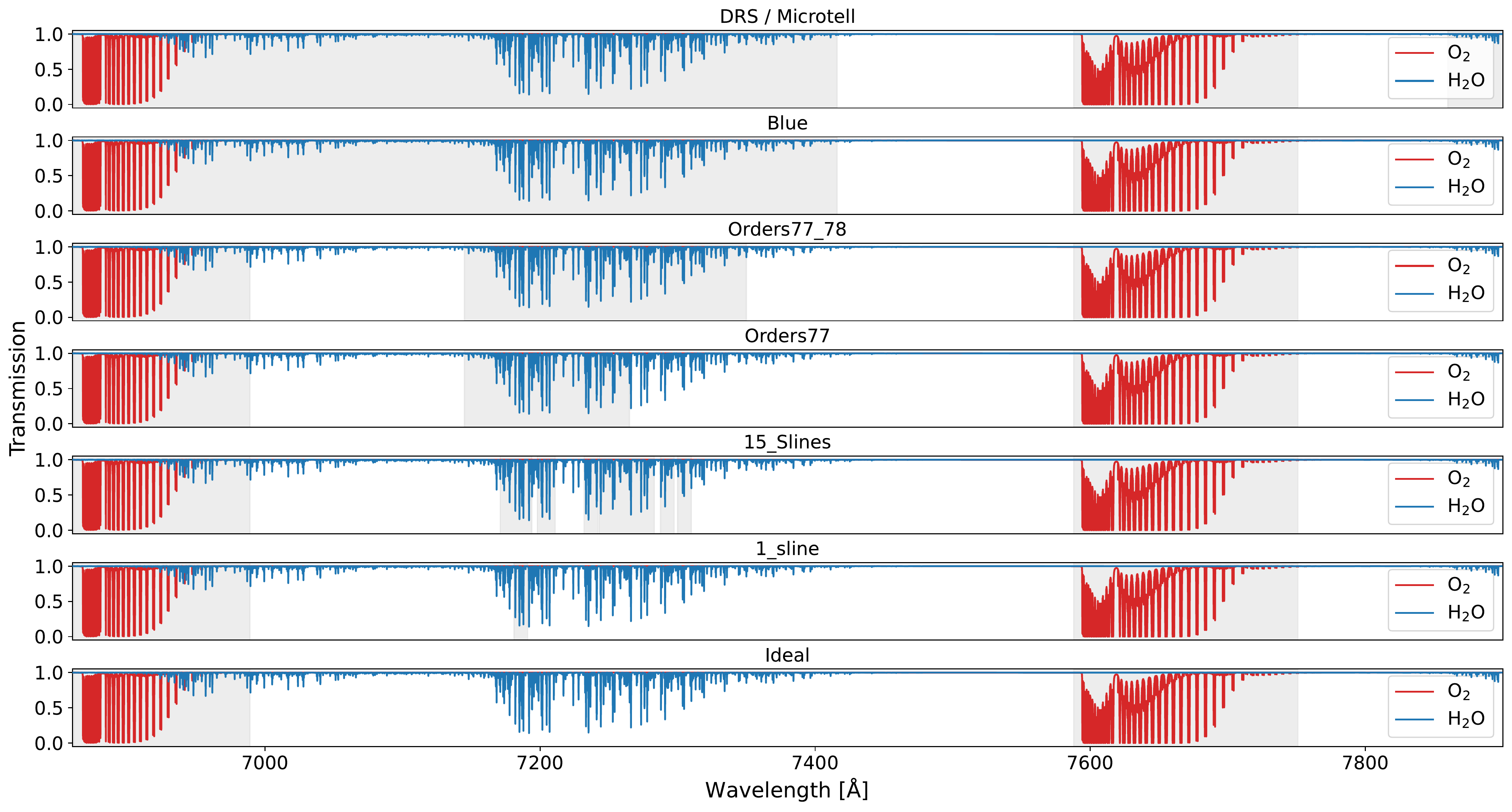}
\adjustbox{minipage=\textwidth,left}{\caption{Inset of Fig.\,\ref{mask_applied} between 6860 and 7900\,\AA.} \label{mask_applied_zoom}}
\end{figure*}

\end{appendix}
\end{document}